\documentclass[preprint,12pt]{elsarticle}

\usepackage{graphicx}   
\usepackage[font=footnotesize]{subfig} 
\usepackage{fixltx2e}
\usepackage{stfloats}
\usepackage{subfloat}
\usepackage{amssymb,amsmath}
\usepackage{url}
\usepackage{array}
\usepackage{appendix}

\newcommand{\mva}{\textsc{\small{MVA}}}
\newcommand{\hess}{\textsc{\small{H.E.S.S.}}}
\newcommand{\hesstwo}{\textsc{\small{HESS-II}}}
\newcommand{\hessend}{\textsc{\small{H.E.S.S\@. }}}
\newcommand{\veritas}{\textsc{\small{VERITAS}}}
\newcommand{\cat}{\textsc{\small{CAT}}}
\newcommand{\gray}{$\gamma$\mbox{-}ray} 
\newcommand{\grays}{$\gamma$\mbox{-}rays} 

\journal{Astroparticle Physics}

\begin{document}
\begin{frontmatter}

\title{A new analysis strategy for detection of faint \gray\ sources with Imaging Atmospheric Cherenkov Telescopes}

\author[APC,LLR]{Y. Becherini\corref{cor1}}
\author[APC]{A. Djannati-Ata\"\i}
\author[APC]{V. Marandon}
\author[APC]{M. Punch} 
\author[APC]{S. Pita}

\cortext[cor1]{corresponding author: Yvonne.Becherini@apc.univ-paris7.fr, Tel: +33 1 57 27 61 58}

\address[APC]{Astroparticule et Cosmologie (APC),
B\^atiment Condorcet, 10, rue Alice Domon et L\'eonie Duquet, F-75205 Paris Cedex 13, France\\
UMR 7164 (CNRS, Universit\'e Paris 7 Denis Diderot, CEA, Observatoire de Paris)}

\address[LLR]{Laboratoire Leprince-Ringuet (LLR), \'Ecole Polytechnique, F-91128 Palaiseau, France\\
UMR 7638 (CNRS, \'Ecole Polytechnique)}



\begin{abstract}

A new background rejection strategy for \gray\ astrophysics 
with stereoscopic Imaging Atmospheric Cherenkov Telescopes (IACT), 
based on Monte Carlo (MC) simulations and 
real background data 
from the \hess\ [High Energy Stereoscopic System, see \cite{hess}.] experiment, 
is described.
The analysis is based on a multivariate combination of both previously-known and 
newly-derived discriminant variables using the physical shower properties, as well as 
its multiple images, for a total of eight variables. 
Two of these new variables are defined thanks to a new energy evaluation procedure, which is also presented here.
The method allows an enhanced sensitivity 
with the current generation of ground-based Cherenkov telescopes to be achieved, 
and at the same time
its main features of rapidity and flexibility allow 
an easy generalization to any type of IACT.
The robustness against Night Sky Background (NSB) variations
of this approach is tested with MC simulated events. 
The overall consistency of the analysis chain has been checked 
by comparison of the real \gray\ signal obtained from \hess\ observations 
with MC simulations and 
through reconstruction of known source spectra. 
Finally, the performance has been evaluated by application to faint \hess\ sources.
The gain in sensitivity as compared to the best standard \textsl{Hillas} analysis ranges 
approximately from $1.2$ to $1.8$ depending on the source characteristics, 
which corresponds to an economy in observation time of a factor $1.4$ to $3.2$.

\end{abstract}

\begin{keyword}
$\gamma$/hadron discrimination; 
weak \gray\ sources; 
Imaging Atmospheric Cherenkov Telescopes; 
multivariate analysis; boosted decision trees.
\end{keyword}

\end{frontmatter}

\section{Introduction}

\label{Intro}

Very High Energy (VHE) \grays\ interact in the Earth's atmosphere, giving rise 
to highly relativistic showers which produce a few-nanosecond flash of Cherenkov photons
in a ``light pool'' of $\sim 250 \; \rm m$ diameter.
This can be detected by one or more ground-based telescopes,
which collect this light using a mirror that reflects 
it onto a camera composed by an array of photomultipliers, 
converting the photons to a charge per pixel,\footnote{\samepage For example, in the first phase of the \hess\ experiment,
the reflective mirror area is $107 \;\rm m^2$ and the pixel size is $0.16^\circ$, 
which is quite well adapted to the shower image size, 
with four such telescopes spaced by $120 \;\rm m$ making up the stereoscopic array, see \cite{crab}.}
and thus giving an image of the shower as seen by each telescope.
\gray\ induced atmospheric showers have a rotational symmetry around the shower axis 
(neglecting geomagnetic effects), 
which results in elliptical images on the cameras.
By combining the information from all the shower images seen,
the arrival direction and the energy of the \gray\ can be reconstructed by well-established methods 
\cite{werner, crab}.
The \gray\ signal is, however, drowned out by an enormous number 
of background air showers initiated by charged hadronic
cosmic-rays. 
Contrarily to the \grays, the hadronic component of air showers is characterized by 
an asymmetric development of the particle cascade due to the hadronic interactions, 
which have large transverse momenta, 
and due to longer-lived muons reaching the telescope altitude.
\par
\vspace{0.4cm}
After decades of research \cite{GRAreview}, 
great progress has been made in the field of VHE \gray\ astronomy in recent years, 
with the advent of the new generation of IACTs. 
The major keys to this success have been stereoscopy -- offering a 
three-dimensional view of the atmospheric shower -- and fast imaging cameras characterized 
both by their time response at the nanosecond scale and by a fine pixel granularity. 
These features have led to larger effective detection areas,
a better angular resolution and a much improved discrimination 
power against the abundant cosmic-ray background.
Nonetheless, 
the detection using standard analysis methods 
of sources at the level of 1\% of the Crab nebula flux requires several tens of hours 
with arrays such as \hess\ or \veritas, even at their optimal sensitivity, out of 
$\rm \sim1000 \; h/year$ of available good-weather moonless nights.
This observation time needed for detection of very low-flux sources, 
for the discovery of new classes of \gray\ emitting sources, 
or for detailed morphological studies of extended sources
can even become prohibitively long if mirror ageing occurs, 
causing an increase in threshold energy and thus a decrease in sensitivity.
\par
\vspace{0.4cm}
Long experience has demonstrated the robustness against changing data-taking conditions
of the standard Hillas-parameter based analysis 
(see \cite{crab} and also a short description in Sec.\ \ref{Stereo}), 
essentially because of its stability under variable conditions, allowing its application 
to the entire set of IACTs world-wide since the 1980's.
However, this kind of analysis 
does not exploit the correlation between images of the same shower as seen by different
telescopes in a simple way, nor take advantage of fine-pixelization imaging.
So, in recent years, additional methods have been developed in order 
to enhance the analysis sensitivity by fitting 
a model of the Cherenkov photons produced by a \gray\ shower to the images seen by each telescope, 
overcoming the above-mentioned limitations.
\par
\vspace{0.4cm}
A first successful development was carried out over a decade ago for the \cat\ experiment 
in mono-telescope mode \cite{LeBohec} with the 2D-model, where the image of the shower 
(in particular its longitudinal profile but also its transverse profile) is used 
to predict the angular origin of the event along the image axis, allowing the event direction on the sky to be obtained, 
as well as giving a $\chi^2$-based discrimination variable for background rejection. 
This method has subsequently been 
improved (passing from a $\chi^2$ minimization to a maximum-likelihood, among other developments), and has been 
optimized to take full advantage of the stereoscopy provided by \hess\ (see \cite{DeNaurois}).
In addition, a three-dimensional (3D-model) reconstruction 
method has been introduced in \hess\ \cite{marianne}, 
where the result of the fit is a set of parameters describing 
the shower characteristics and 
where the selection analysis cuts are defined on the physical shower properties, rather than on the likelihood value.
An upgraded version of the 3D-model analysis introduced a cut based on a $\chi^2$ value in order to 
include the information from 
untriggered telescopes, see \cite{Melitta}.
More recently, two different types of multivariate analyses (\mva) have been applied for \hess: 
the first using Hillas-parameter based discriminant variables \cite{ohm}; the second using discrimination 
based on Hillas-parameter based variables, 
plus 3D-model parameters and 2D-model goodness-of-fit \cite{Fiasson}.
\par
\vspace{0.4cm}
The new method presented in this paper is 
based on an \mva\ approach optimized for the detection of low-flux sources,
in which new ideas have also been implemented. 
First of all, the decision was taken to explore only discriminant parameters 
which are not a goodness-of-fit value, principally so as to lower the sensitivity 
of the analysis to uncertainties due to NSB variations. Indeed, 
in order for the systematics due to the NSB to be under control, 
goodness-of-fit based discrimination analyses need a 
rather precise knowledge of the underlying probability distribution
(for likelihood fits)\footnote{Or its error in case of a $\chi^2$ fit.}  
of the pixel response to \gray\ showers and noise,
together with its variation as a function of the NSB rate 
(see for example Sec.\ 3.7 of \cite{DeNaurois} and the $\chi^2$ definition of \cite{LeBohec}). 
These distributions may also be sensitive to imperfections of 
the electronics or the acquisition system and their drift over time, 
which might be difficult to reproduce\footnote{Some years of operation 
experience have been found to be necessary to understand and characterize such effects.} 
or time-costly to implement with MC simulations of the instrument. 
Avoiding such types of parameters should provide a more ``portable'' method, applicable 
to many current and future IACTs.
Secondly, the information carried by the 3D-model about the predicted shape of \gray\ images 
on the cameras is found to provide additional discrimination power
when the Hillas moments of the predicted images are calculated and used to derive new 
discriminant variables. 
Thirdly, an energy reconstruction method has been developed explicitly for this work, 
the conception of which has allowed the definition of two new discriminant parameters. 
These parameters, combined with a third new one -- all of which are detailed below --
provide additional power for hadron rejection.
Special attention has been paid in this work to the cut optimization for different expected source characteristics 
(intensity and spectral hardness), as well as for studies of source morphology.
\par
\vspace{0.4cm}
After a brief overview of the choices made for stereo reconstruction (Sec.\ \ref{Stereo})
a versatile 
\gray\ energy reconstruction algorithm will be presented in Sec.\ \ref{OakEnergy}, 
which is used as the basis for two newly-developed parameters.
The parameters chosen for the $\gamma$/hadron discrimination are introduced in Sec.\ \ref{SelePara}, while 
the details of the multivariate event classification method are presented in Sec.\ \ref{MVA}.
The stability of the method is tested through comparison of MC simulations and real \gray\ data 
in Sec.\ \ref{Robu}. Finally, in Sec.\ \ref{Perf} the performance is evaluated through application to \hess\ sources.

\section{Stereo reconstruction}

\label{Stereo}
The steps of the shower reconstruction process used for the present work are described here. 
First of all, the images triggered by the event in the telescopes 
are cleaned so as to eliminate the pixels not containing a significant fraction of the Cherenkov photon signal
(comparable to the NSB fluctuations) using 
the methodology described in \cite{crab}\footnote{
The two-threshold image cleaning used in this work 
is of type 5--$10\;\rm p.e.$; and a pixel is also rejected if its charge 
is less than $3 \sigma$ above its pedestal level. 
This is the case for the standard \hess\ analysis. 
Other IACTs use thresholds which are adapted to their instruments.}.
The charges of the selected pixels are used to calculate the first and second moments of the images in each telescope, 
allowing the ``Hillas parameters'' to be derived (equivalent to fitting an ellipse to each image).
A Hillas ellipse is characterized by its {\textsl{length}} and {\textsl{width}},  
as well as by its position with respect to the camera centre, 
by the total charge in photo-electrons (p.e.), and by its main axis direction.
Images whose centres-of-gravity fall near the edge of the cameras are discarded so as to minimize 
shower parameter misreconstruction due to image truncation; 
all the remaining images passing a pre-defined charge threshold 
are used to calculate the shower arrival direction and 
the intersection of the shower axis with the ground (also known as \textsl{impact position}), see \cite{werner, crab}.
The geometry of the stereoscopic images allows the slant depth of the shower maximum to be estimated, 
assuming the centre of gravity of the images to be the projection of the maximum into the image plane.
\par
\vspace{0.4cm}
The shower characteristics derived from the Hillas-parameter based variables 
are useful as such, but also as the first guess for the subsequent 3D-model optimization \cite{marianne}.  
In the 3D-model, the \gray\ shower is modelled by a 3D-photosphere 
of Cherenkov photon origins in the atmosphere which is assumed to have a Gaussian distribution along all axes. 
This photosphere is described by eight parameters which are intrinsic to the shower: 
the direction and the impact position of the incident \gray, 
the photosphere length (along the shower axis) 
and transverse width (rotational symmetry around the shower axis being assumed), 
the slant depth of the shower maximum, 
and the number of Cherenkov photons generated by the electrons and positrons in the shower development. 
The 3D-model is used to predict the distribution of Cherenkov 
light in the cameras of a telescope array; then 
the intrinsic parameters of the shower are adjusted to achieve a good fit 
(using a maximum-likelihood optimization) of the ensemble of \gray\ shower images.
\par
\vspace{0.4cm}
Given the Hillas and the 3D-model reconstruction procedures, two estimations 
of the event direction are available. 
The 3D-model point-spread-function (PSF) is better than that provided by the Hillas-parameter based reconstruction 
for energies below $500 \;\rm GeV$, 
whereas the inverse applies for higher energies. 
In this work, the Hillas-parameter based direction has been chosen for simplicity of implementation.
Studies of the use of optimized combination of the two reconstructed directions will be undertaken for a future upgrade.

\section{Energy reconstruction}
\label{OakEnergy}

In ground-based \gray\ astronomy,
sources are observed under various conditions of zenith angle, under which the images produced 
by \gray\ showers propagating in the atmosphere vary markedly, and 
the shower images also vary depending on the impact parameter of the shower with respect 
to the given telescope position
(see for example the two-dimensional image profiles for such varying conditions in \cite{Buckley}).  
In addition, sources may be observed at different offsets with respect 
to the centre of the camera, resulting in different detector responses.
Moreover, any loss of photon collection or detection efficiency 
can play an important role in the evaluation 
of the total image charge amplitude.
Therefore, the data analysis in this field is characterised 
by a wide use of lookup tables, in which the \gray\ image characteristics 
as a function of the varying data taking conditions are stored 
to allow the modelling of the detector response to signal events.
\par
\vspace{0.4cm}
The energy of the \gray\ shower can be reconstructed by using such  
lookup tables generated either with the Hillas-parameter based information, or with that provided by the 3D-model. 
However, the method for the evaluation of the energy with the 3D-model requires 
the convergence of the model fit, which is not guaranteed, especially for low-charge two-telescope events.
As a generic method usable for all analysis configurations was deemed preferable, 
the choice was made to work in the Hillas-parameter based reconstruction framework.
The proposed method of calculating the \gray\ energy
has the advantage of being flexible, because the definition of the charge profiles 
can be performed with any kind of \gray\ simulation, 
and moreover, the energy evaluation phase is relatively fast with respect to other 
alternative methods, such as those requiring minimization \cite{Lemiere}.
The energy reconstruction method presented here is composed of two separate phases: a preparatory phase 
and an evaluation phase. 

\subsection{Preparatory phase}
\label{Prepa}
The preparatory phase requires the generation of lookup tables 
(\textsl{charge profiles}, see for example Fig.\ \ref{Oak}a) 
of the sum of the charges measured in a simulated \gray\ cleaned image as a function of the impact parameter 
(referred to below as $R$). 
These are created for discrete sets of values of the simulated energy $E_{\rm true}$, 
of zenith angles $\ensuremath{\Theta}$, of values of the mirror efficiency $M$, 
of offsets $D$ of the shower angular origin with respect to the centre of the camera, 
and for multiple bands of reconstructed shower maximum depths $H$
(as estimated from the Hillas parameters of the images). 
Note that these charge profiles are related to 
the Cherenkov light profile on the ground, but as detected in the images seen by the telescopes 
(including the photomultiplier efficiency with wavelength and image cleaning effects, etc.).

\subsection{Energy evaluation phase}
\label{Eval}
\begin{figure}[t]
  \centerline{  
    \subfloat[]{\includegraphics[width=7cm]{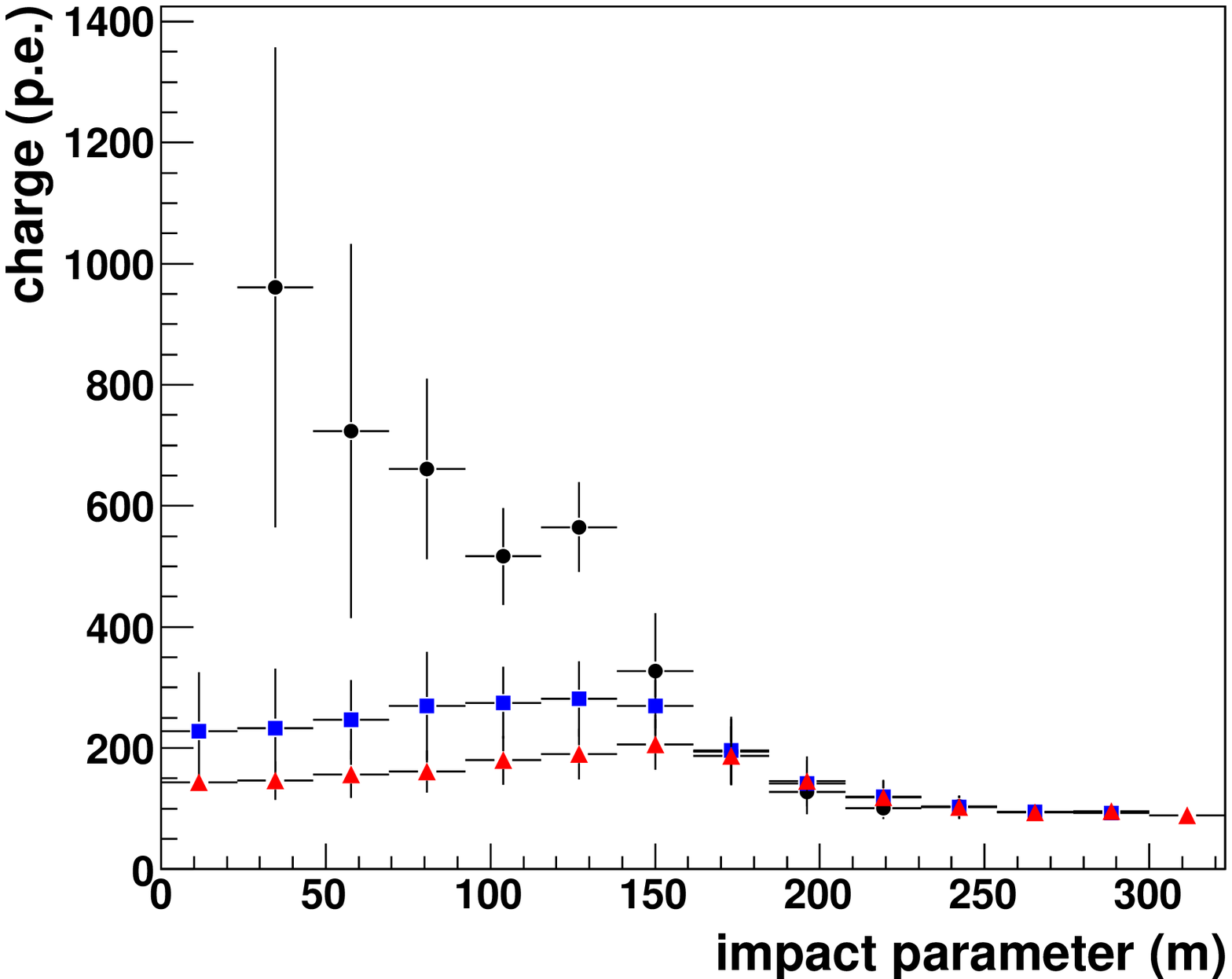}}
    \hspace{1cm}
    \subfloat[]{\includegraphics[width=6cm]{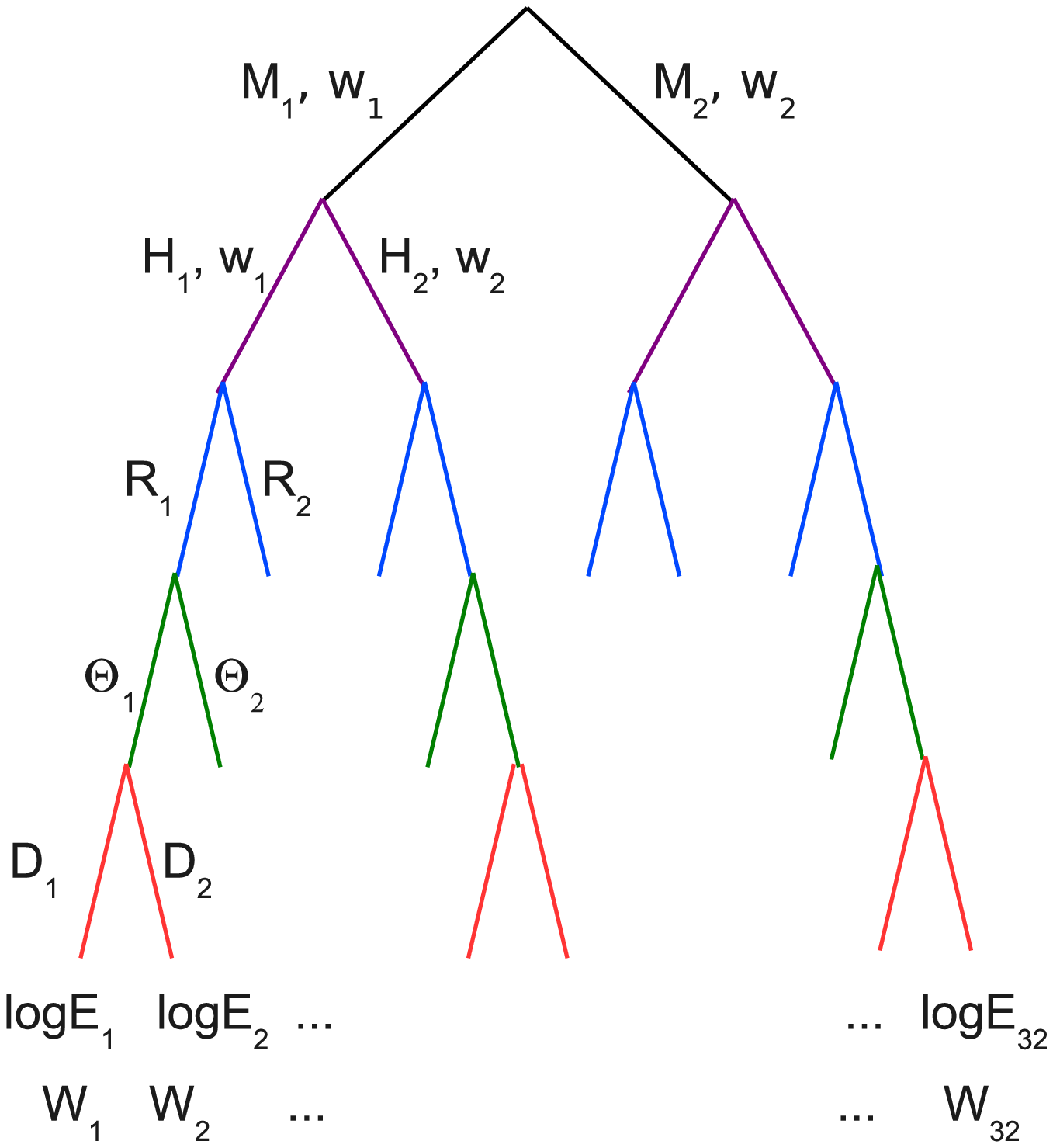}}
  }
  \caption{\small{(a) Profiles of the charges of the cleaned images in the cameras as a function 
      of the reconstructed impact parameters at three different observing angles 
      $0^\circ$ (points), $32^\circ$ (squares) and $41^\circ$ (triangles), 
      for a given energy, shower maximum, offset and mirror efficiency; 
      the error bars shown represent the spread of the distributions.
      (b) Schematic of the tree used for the evaluation of the event energy for each telescope.
      For an event having a zenith angle 
      $\ensuremath{\Theta}_{\rm reco} \in [ \ensuremath{\Theta}_1, \ensuremath{\Theta}_2 ]$, 
      an offset $D_{\rm reco} \in [D_1, D_2]$, 
      a maximum depth $H_{\rm reco} \in [H_1,H_2]$, an impact parameter $R_{\rm reco}\in [R_1, R_2]$, 
      and being observed in a telescope having a mean mirror efficiency $M_{{\rm Tel}_i} \in [M_1, M_2]$,
      the $\log_{10}E_{j}$ and the weight $W_{j}$ are calculated for 
      each branch of the tree and then the different values are combined to evaluate 
      the event energy per telescope $E_{{\rm Tel}_i}$.
  }}
  \label{Oak}
\end{figure}

\begin{figure}[t]
  {\subfloat[]  {\includegraphics[width=3in]{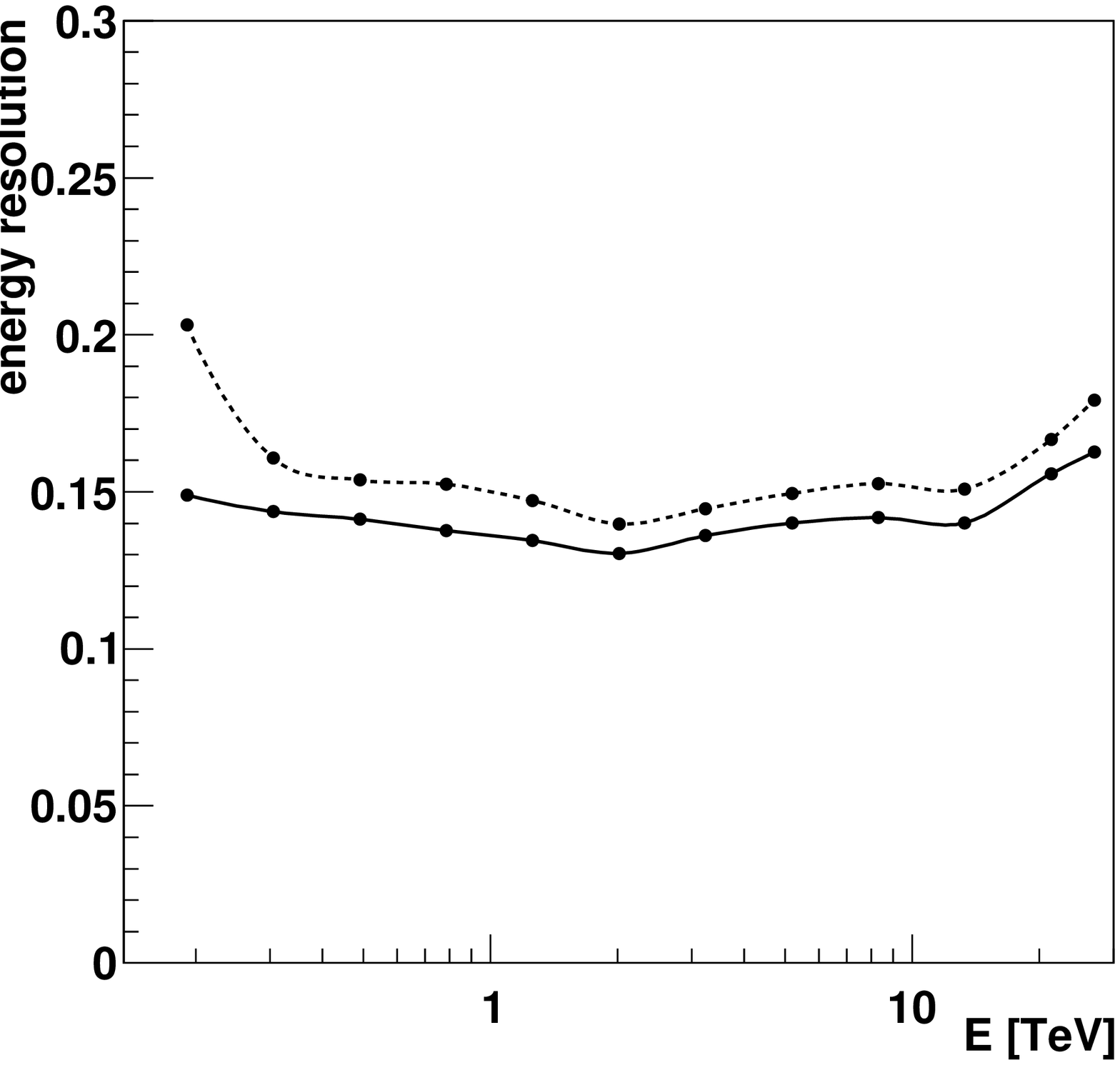}}}
  {\subfloat[]  {\includegraphics[width=3in]{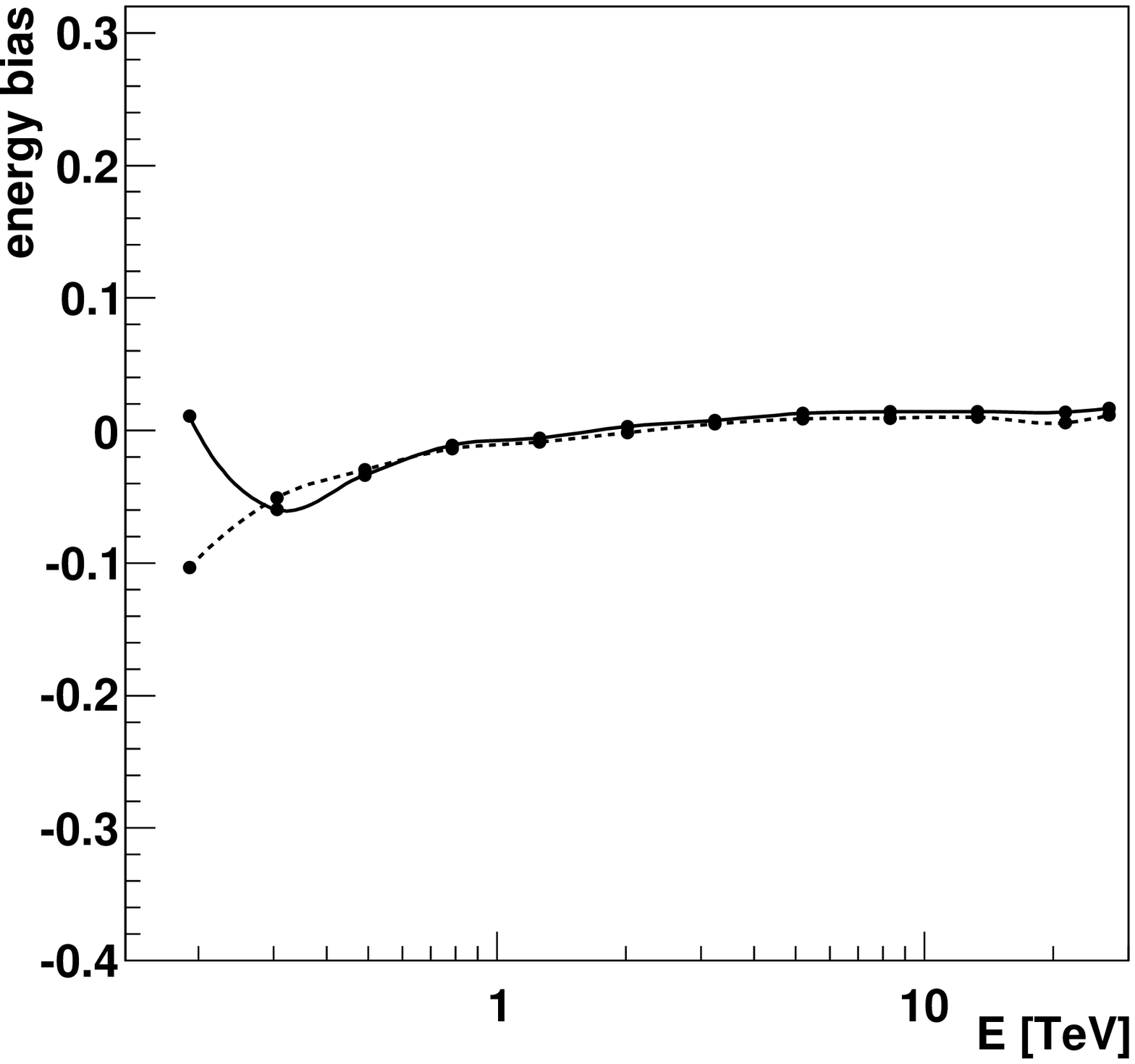}}}
  \caption{\small{
      Given the distribution of $\log(E_{\rm reco}/E_{\rm true})$, where $E_{\rm reco}$ 
      and $E_{\rm true}$ are the reconstructed and simulated energy values,
      its standard deviation defines the 
      energy resolution (a), and the mean value
      defines the bias (b).
      The two curves refer to two analysis configurations 
      ($40\;\rm p.e.$ and $150\;\rm p.e.$, dashed and continuous line respectively, see Sec.\ \ref{HardSoftSpectra}). 
  }}
  \label{EnergyResol}
\end{figure}

The energy evaluation phase consists of the use of these pre-generated charge profiles 
to calculate the event energy based on a weighted combination 
of the estimates of the energy in each telescope.
The strategy for the evaluation of the energy in each telescope can be easily visualized by a tree, 
see Fig.\ \ref{Oak}b.
For an event having a triggered telescope ${\rm Tel}_{i}$ with an associated 
mirror efficiency $M_{{\rm Tel}_{i}}$ and a measured charge $q_{{\rm Tel}_{i}}$, 
there are four reconstructed parameters:
the zenith angle $\ensuremath{\Theta_{\rm reco}}$, the offset $D_{\rm reco}$, 
the maximum depth $H_{\rm reco}$ and the impact parameter 
with respect to the position of the telescope $R_{\rm reco}$.
The weight of each branch of the tree $W_{j}$ can be calculated, where $W_{j}$ is given by:
\begin{equation}
  W_{j} =  \prod_{k=1}^{5} {p_{k}} \; \; {\rm where} \; \; p_k = w_k \; \; {\rm or} \;  p_k = (1-w_k) 
  \rm \; depending \; on \; the \; branch \; taken, 
  \label{Wj}
\end{equation}
and where the weights $w_{k}$ are given by the interpolation of the nearest simulated values. 
As an example, for an event having a reconstructed zenith angle $\ensuremath{\Theta_{\rm reco}}$ the algorithm searches 
for the two fixed values $\ensuremath{[ \Theta_1, \Theta_2 ]}$ between which to interpolate such that 
$\ensuremath{\Theta}_{\rm reco} \in [ \ensuremath{\Theta}_1, \ensuremath{\Theta}_2 ]$, 
and applies a weight of 
${w_\ensuremath{\Theta} = (\ensuremath{\Theta}_{\rm reco}-\ensuremath{\Theta}_1)/(\ensuremath{\Theta}_2-\ensuremath{\Theta}_1)}$.
For each branch of the tree, given the value of the measured charge ${q_{{\rm Tel}_{i}}}$, 
the event energy associated with the telescope can be retrieved using the charge profiles mentioned 
in Sec.\ \ref{Prepa}, which were defined for a set of energy values.  
The energy in each triggered telescope can thus be easily calculated 
by a weighted mean on all the energy values and weights obtained from the $2^5 = 32$ branches:
\begin{equation}
  \log_{10}E_{{\rm Tel}_{i}} = \frac{\sum_{j=1}^{32} \left[{W_{j}\cdot \log_{10} E_{j}}\right]}{\sum_{j=1}^{32} {W_{j}}}
  \label{ETel}
\end{equation}
The energies $E_{{\rm Tel}_i}$ calculated separately for each triggered telescope  
are combined in the final step with a charge-weighted mean:
\begin{equation}
  E_{\rm reco} = \frac{1}{Q_{\rm tot}} \sum_{i=1}^{N_{\rm Tels}} \left[{q_{{\rm Tel}_{i}}\cdot E_{{\rm Tel}_{i}}}\right],\;
  {\rm where}\; {Q_{\rm tot}} = \sum_{i=1}^{N_{\rm Tels}} {q_{{\rm Tel}_{i}}}
  \label{MeanE}
\end{equation}
$Q_{\rm{tot}}$ being the total charge of the event.
The resolution and bias of the current implementation of the algorithm can be seen
in Fig.\ \ref{EnergyResol}: 
in the energy range 0.2--$30\;\rm TeV$ the resolution is almost stable at $\sim 15\%$,  
while the bias is better than $5\%$ for $E>0.3 \;\rm TeV$.

\section{Selected $\gamma$/hadron discriminant parameters}
\label{SelePara}

The definitions of the discriminant parameters 
take advantage of the information carried by the two different
stereo reconstruction algorithms discussed in Sec.\ \ref{Stereo} (the 3D-model and Hillas-parameter based ones)
and of the energy reconstruction algorithm introduced in Sec.\ \ref{Eval}. 
As will be mentioned in Sec.\ \ref{importance}, the parameters have been chosen 
according to their discriminant power and in order to avoid strongly-correlated variables.

 \begin{figure}[t]
  \subfloat[]{\includegraphics[width=2.1in]{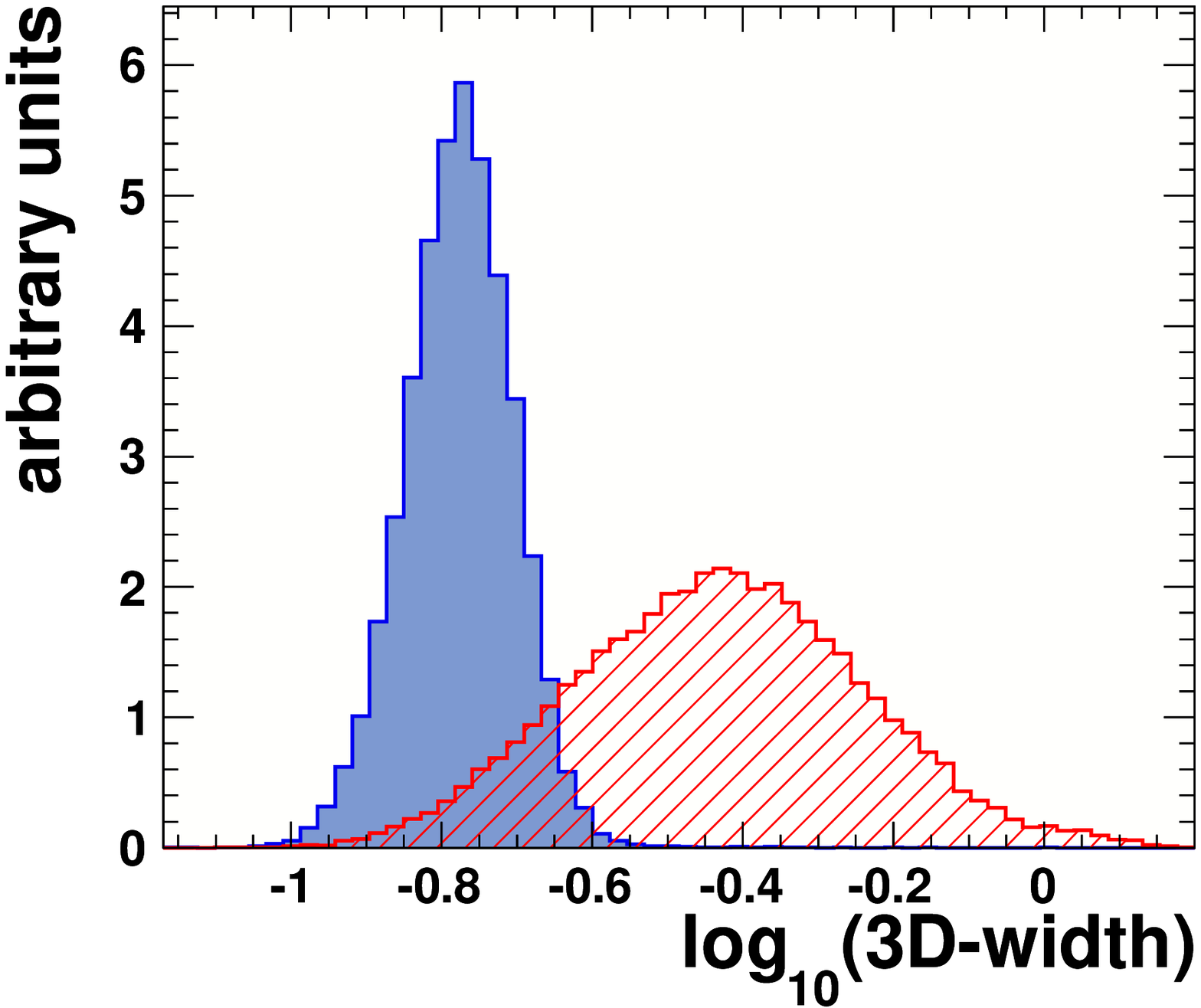}}
  \subfloat[]{\includegraphics[width=2.1in]{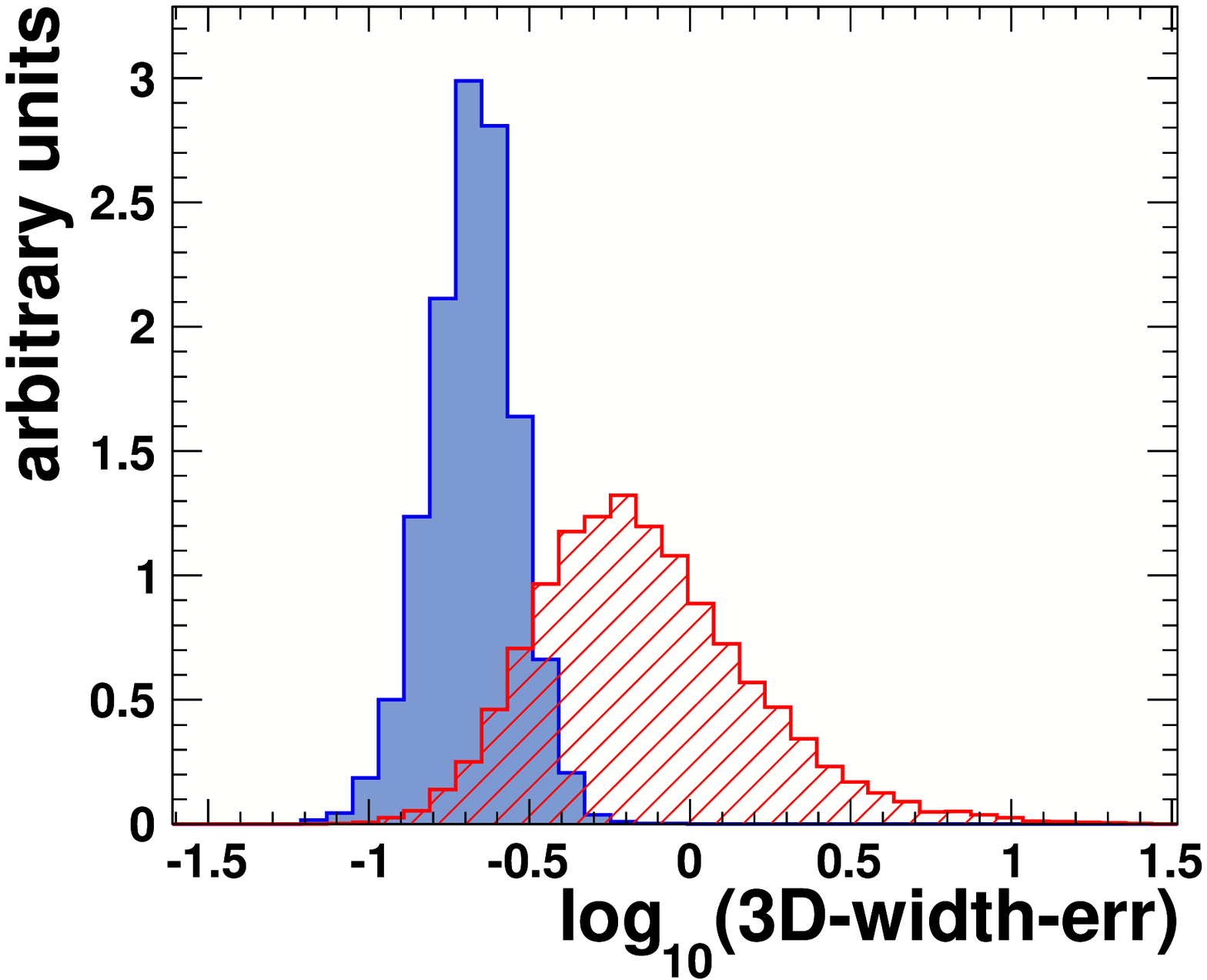}}
  \subfloat[]{\includegraphics[width=2.1in]{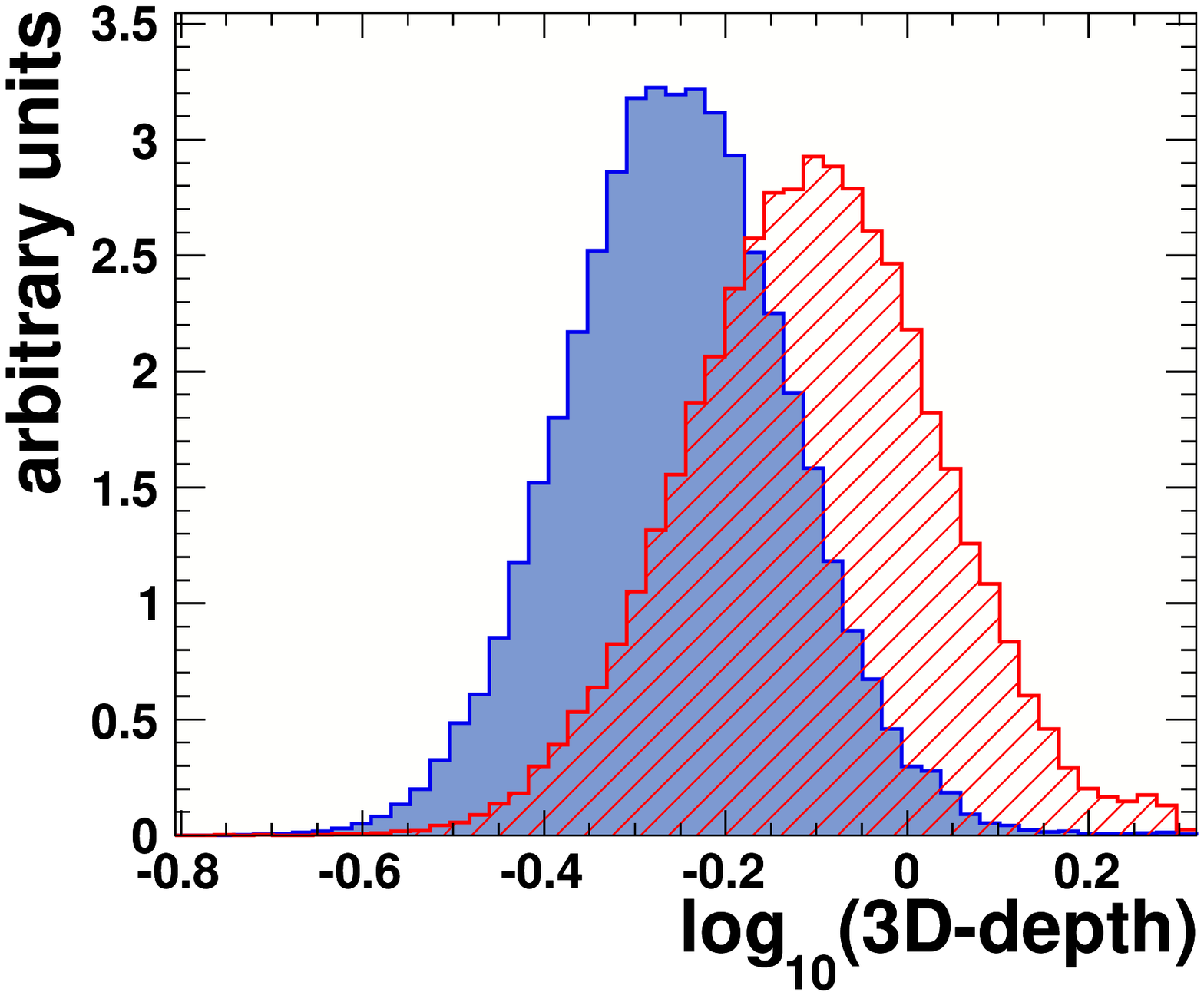}}
  \caption{\small{3D-width (a) 
      and 3D-width-err (b) for the \hess\ array and for energies in the range 1--$2\;\rm TeV$. 
      The 3D-depth (c) is shown for the range 100--$300\;\rm GeV$. 
      All three parameters are shown for the $10$--$20^\circ$ zenith angle range.
      Filled and hatched distributions represent the \gray\ simulated events and the measured 
      cosmic-ray background, respectively.}}
  \label{Model3DPar}
\end{figure}

\subsection{Hillas-parameter based discriminant variables}
\label{HillasPara}

For a given reconstructed shower, two powerful discriminant parameters 
commonly used for stereoscopic arrays
can be defined: 
a \textsl{Mean Scaled Length} (MSCL) and a \textsl{Mean Scaled Width} (MSCW).
As described in \cite{crab}, lookup tables are prepared based on simulations, which contain the mean width and length 
and their respective rms scatter for a \gray\ as a function of the amplitude of the shower image in the camera 
and the impact parameter relative to the telescope.
The MSCL and MSCW are then calculated as the mean values of the deviations of the measured lengths and widths
from the predicted mean values, divided by their rms. 
In short, these mean scaled parameters indicate deviations of the shape of the detected images 
compared to those resulting from the \gray\ simulations, and hence allow for discrimination, 
see Fig.\ 7 of \cite{crab} for further details.
The Hillas-parameter based reconstruction and discrimination have been the basis of a large number 
of discoveries in recent years, showing that these two simple scaled parameters were sufficient 
for an efficient $\gamma$/hadron discrimination in case of bright sources 
(i.e., $> 10\%$ C.U.\footnote{``Crab Units'', 
i.e., compared to a hypothetical source with the Crab nebula intensity and spectrum at the 
zenith angle of observation.}).  

\subsection{3D-model based discriminant parameters}
\label{ModelPara}

As described in Sec.\ \ref{Stereo}, eight parameters are adjusted by the 3D-model fit, 
which also provides the error estimates for these parameters.
Among these, two parameters, 
the \textsl{reduced 3D-width}\footnote{
As the Cherenkov photosphere is found to be dependent 
on the variation of the Cherenkov threshold with the
altitude of the shower maximum, an efficient and dimensionless 
$\gamma$/hadron discrimination parameter called \textsl{reduced
3D-width} has been defined, though for simplicity this is referred to as the 3D-width in the following.} 
and the depth of shower maximum (\textsl{3D-depth}), 
together with the error on the former parameter (\textsl{3D-width-err}), 
have been retained for their high discriminant power.
The distribution for the 3D-width for \grays\ (Fig.\ \ref{Model3DPar}a) 
is narrower and has a peak at lower values compared to that of the background.
The same applies to the 3D-width-err (Fig.\ \ref{Model3DPar}b), 
showing that the model is not suitable for the representation of a hadron-induced shower, 
thus giving larger error values as a consequence, with an associated discrimination power.
Moreover, it was found that the two 
3D-width related parameters show a significant separation performance over the entire energy range, while 
3D-depth (Fig.\ \ref{Model3DPar}c)
is highly discriminant for lower energies, with decreasing separation power with increasing energy, 
as will be shown also in Sec.\ \ref{importance}.
\begin{figure}[t]
  \centerline{
    {\includegraphics[width=3in]{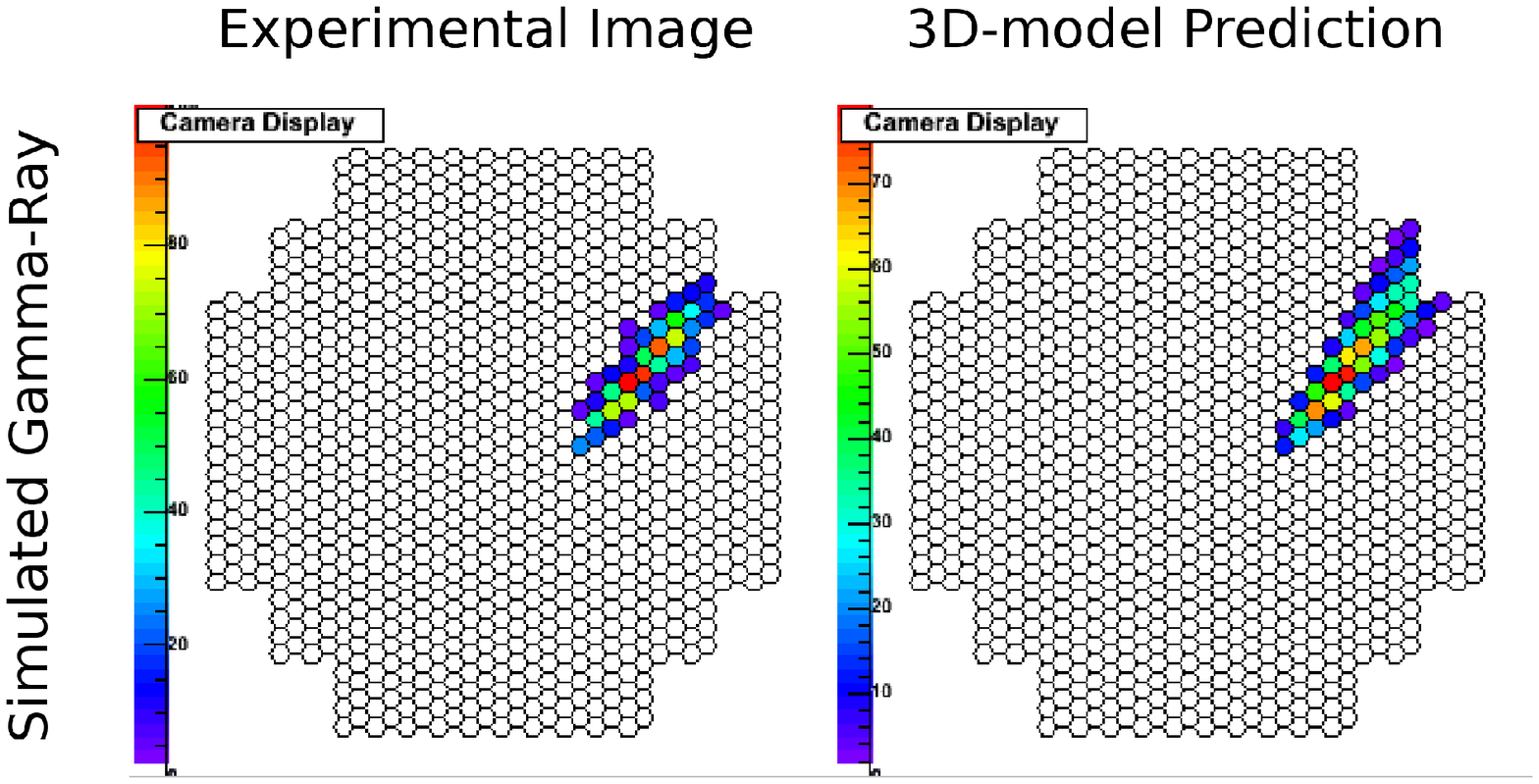}}}
  \centerline{
    {\includegraphics[width=3in]{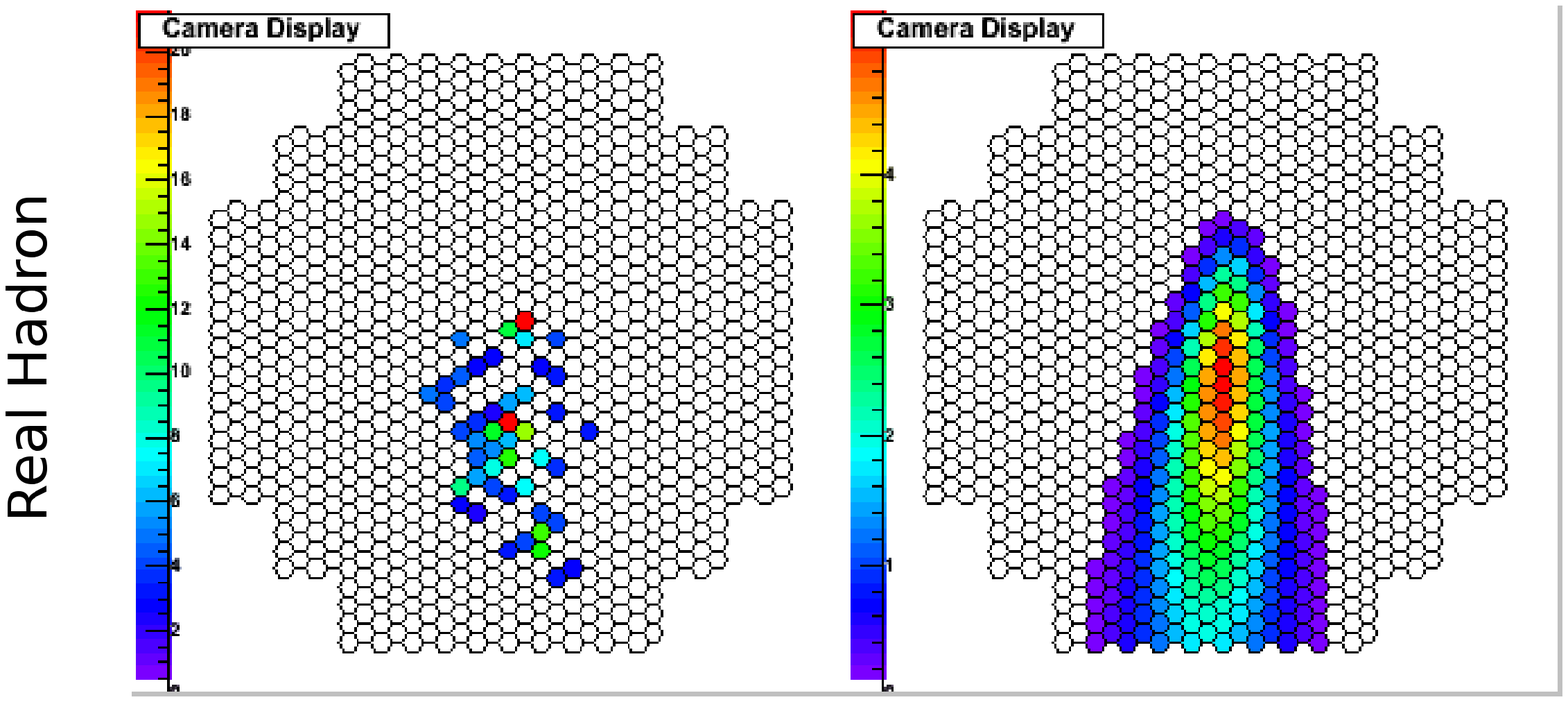}}}
  \caption{\small{
      \gray\ and hadron images seen in a camera.
      The upper panels show the simulated \gray\ cleaned image (\textsl{left panel}) 
      and corresponding prediction given by the 3D-model optimization (\textsl{right panel}), while
      the lower panels show a real cleaned hadron image and its corresponding 3D-model prediction.}}
  \label{Model}
\end{figure}

\subsection{New discriminant parameters}
\label{NewPara}

Beyond the five above-mentioned discriminant parameters, 
three new ones have been defined for the purpose 
of further enhancing the $\gamma$/hadron separation performance.
Due to the difference in the shower development, especially regarding
its rotational asymmetry/symmetry around the shower axis, even in the case where the fit of a hadronic
shower with a \gray\ model (as for instance in \cite{marianne}) succeeds,
it may lead to incoherencies -- in both shape and total charge -- between the predicted shower images
and those observed (or absent) in the individual telescopes  (see Fig.\ \ref{Model}).
The alternative method proposed here is to explore these differences by defining 
additional Hillas parameters  (here called \textsl{HillasOnModel} parameters) 
using the predicted images in the cameras, 
to which the same cleaning and moment calculation procedures are applied and 
which can be used to define new discriminant variables.
In this way the inter-telescope correlations which are inherently included by the 3D-model optimization 
are further exploited in a simple Hillas-parameter based discrimination framework. 

\subsubsection{The $\Omega$ parameter}

\begin{figure}[t]
  \subfloat[]{\includegraphics[width=2.1in]{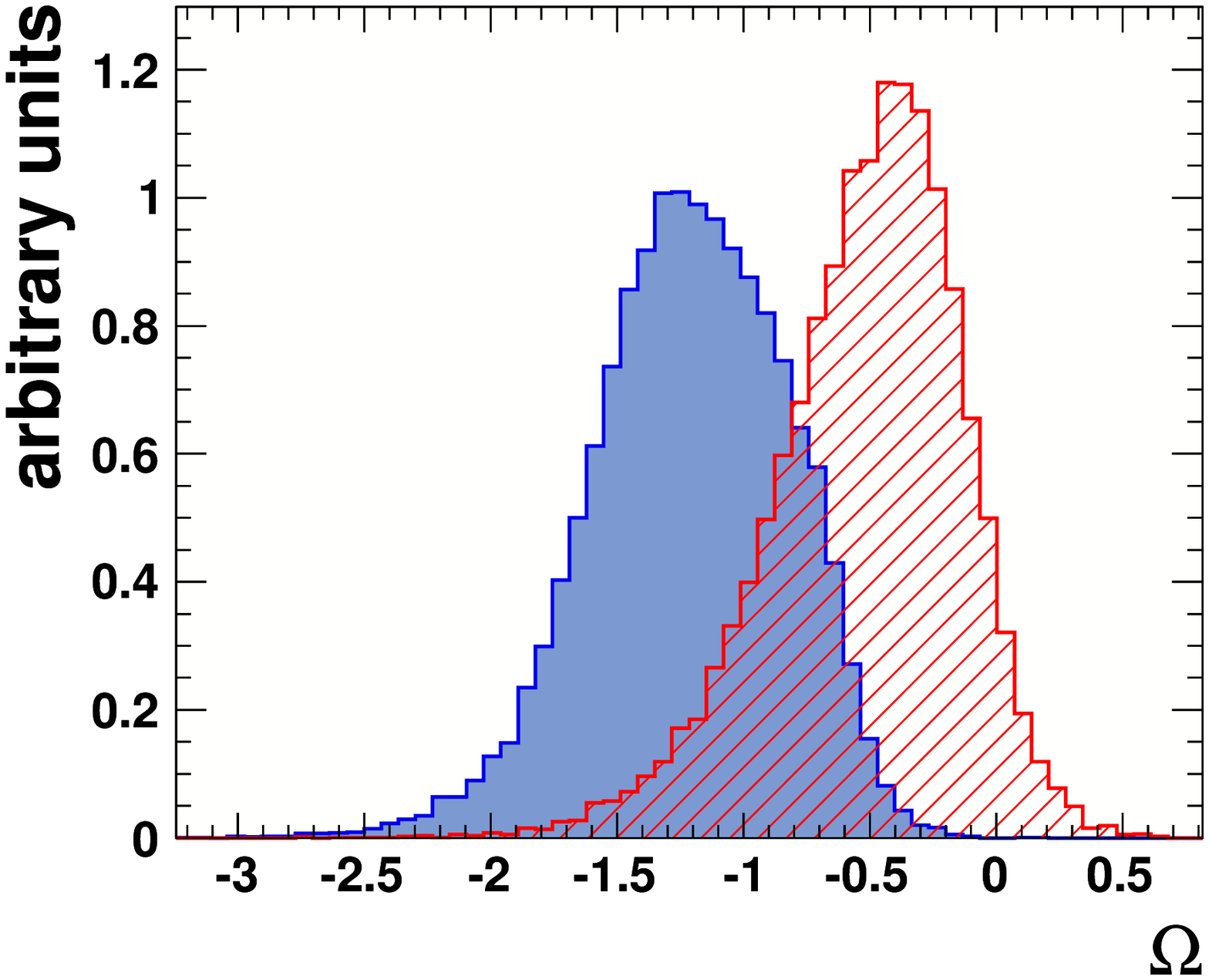}}
  \subfloat[]{\includegraphics[width=2.1in]{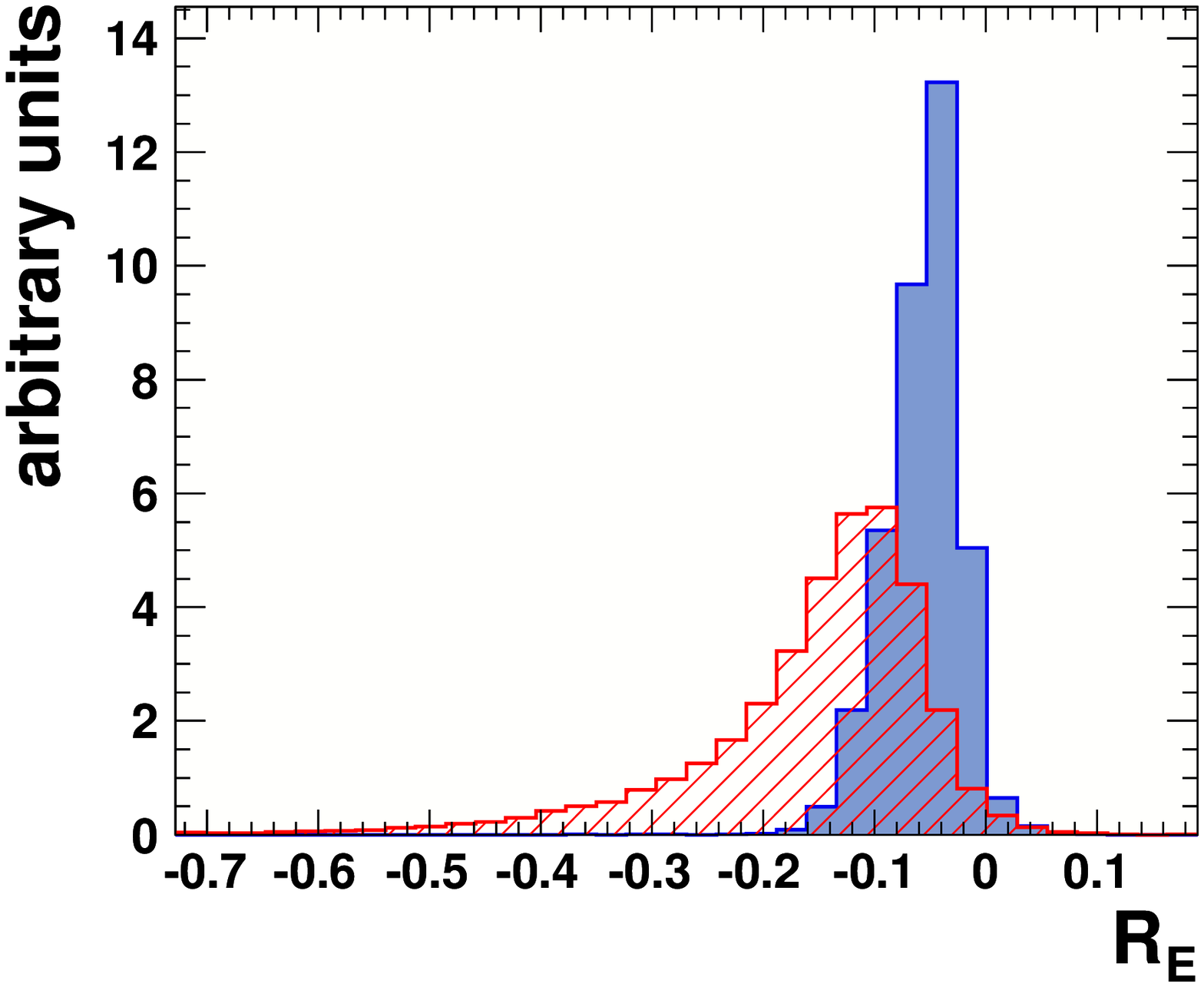}}
  \subfloat[]{\includegraphics[width=2.1in]{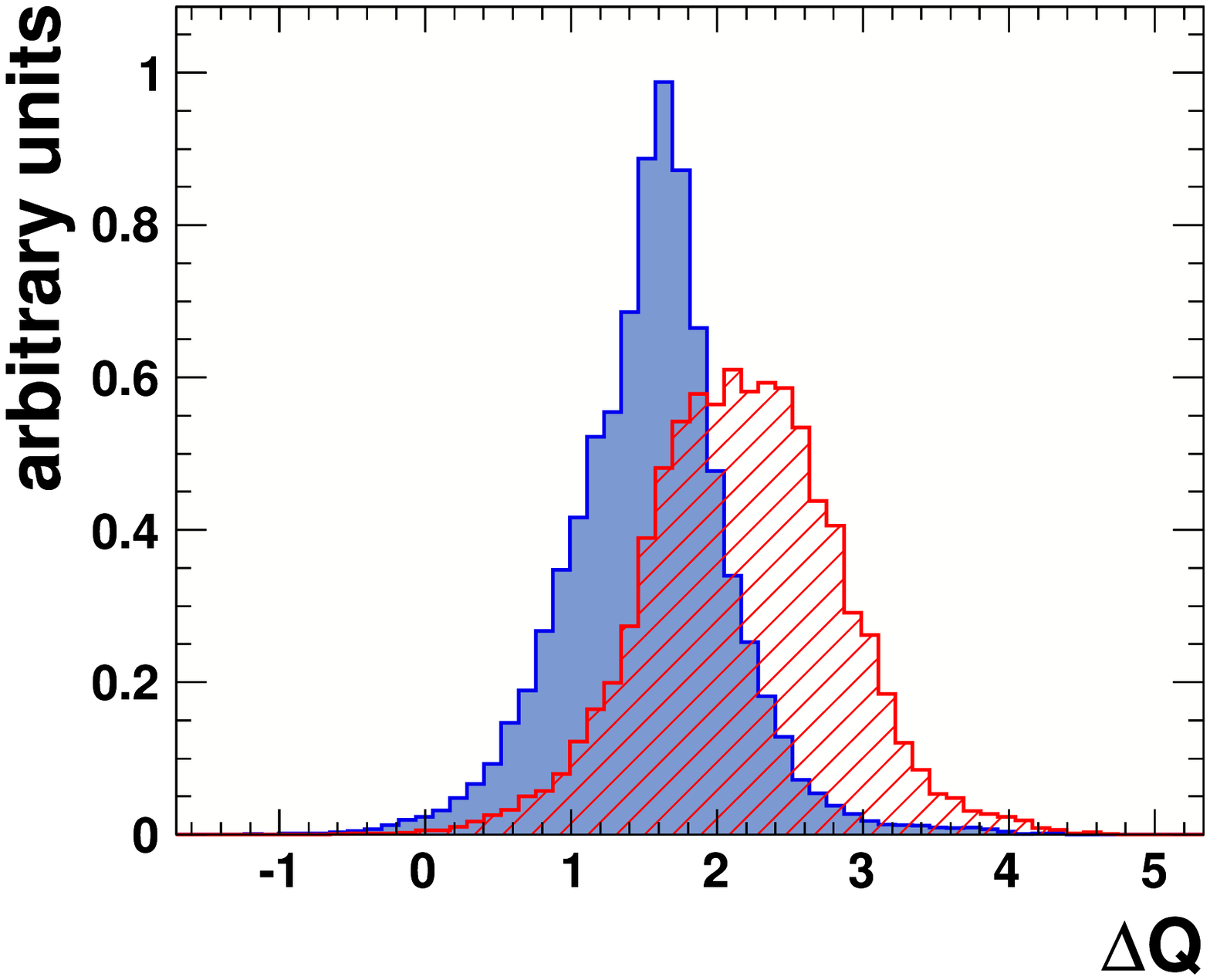}}
  \caption{\small{The $\Omega$ (a), $R_{E}$ (b) and $\Delta Q$ (c) distributions for the energy range 1--$2\;\rm TeV$ 
      and for the zenith angle range $10$--$20^\circ$.
      The filled distributions correspond to the \gray\ events, while the hatched distributions 
      correspond to the cosmic-ray background.}}
  \label{NewPars}
\end{figure}

The discriminant parameter named $\Omega$ is based on the expected difference in the angular distance on the sky 
between the two reconstructed shower directions when using  
\textsl{Hillas} and the \textsl{Hillas\-On\-Model} ellipses:
\begin{equation}
  \Omega = \log_{10}(\arccos(\hat{v}_{\rm Hillas}\cdot \hat{v}_{\rm HillasOnModel}))
  \label{omega}
\end{equation}
where $\hat{v}_{\rm Hillas}$ and $\hat{v}_{\rm HillasOnModel}$ are the normalized vectors 
corresponding to the two directions. 
The shift in the mean value of the $\Omega$ distributions  (see
Fig.\ \ref{NewPars}a) is due to the fact that the difference in the major axis of the
two ellipses is small for a well-fitted \gray\
and larger for a badly-fitted hadron.

\subsubsection{The $R_{E}$ parameter}
 
The energy reconstruction procedure described in Sec.\ \ref{OakEnergy} allows 
the derivation of the following parameter based on the ratio of two reconstructed energy values: 
\begin{equation}
  R_{E} = \log_{10} \left[\frac{E_{\rm Hillas}}{E_{\rm HillasOnModel}}\right]
  \label{DE}
\end{equation}
where the energy of the event, $E_{\rm Hillas}$, is estimated with the Hillas parameters of the detected images, 
and the additional energy estimation, $E_{\rm HillasOnModel}$, is carried out based on 
the \textsl{HillasOnModel} parameters.
This ratio is expected to have
small deviations from zero for well-reconstructed \grays, 
while for misreconstructed hadrons larger deviations are expected.
The distributions of $R_{E}$ for simulated \grays\ and 
for real background are illustrated in Fig.\ \ref{NewPars}b, 
and these differ as expected, 
the background having a large tail towards smaller $R_{E}$.

\subsubsection{The $\Delta Q$ parameter}

The third new parameter also takes advantage of the new energy reconstruction procedure. 
Having calculated the event energy and knowing the reconstructed impact parameter of the event, 
the second phase of the procedure for the energy evaluation per telescope described in Sec.\ \ref{OakEnergy} 
can be inverted to find the expected image charges $\tilde{q}_{{\rm Tel}_{i}}$
on each telescope. 
A discriminant parameter based on the deviation of $\tilde{q}_{{\rm Tel}_{i}}$ 
with respect to the 
measured charges $q_{{\rm Tel}_i}$ can be defined using a least squares in charge:
\begin{equation}
  \Delta Q = \frac{1}{Q_{\rm tot}} \sum_{i=1}^{N_{\rm Tels}} \left[{\tilde{q}_{{\rm Tel}_i} - q_{{\rm Tel}_i}}\right]^2
  \label{Missing}
\end{equation}
where $N_{\rm{Tels}}$ is the number of telescopes in the array and $Q_{\rm tot}$ is the summed charge in all telescopes
as defined previously in Eq.\ \ref{MeanE}.
For poorly-reconstructed events, an inconsistency between the energy evaluation and the reconstructed impact position 
is expected.
Hence, the predicted charge $\tilde{q}_{{\rm Tel}_i}$ on a telescope for such an event is expected to  
be very different from the $q_{{\rm Tel}_i}$ measured 
(or not measured, in the case where a telescope did not trigger). 
As shown in Fig.\ \ref{NewPars}c, the shape of the $\Delta Q$ distributions reflect the expected behaviour, 
the background events having larger values.  

\section{Multivariate analysis} 

\label{MVA}
In order to achieve an efficient hadron rejection, all the above-mentioned parameters
are used within a multivariate discrimination scheme based on the \textsc{\small{TMVA}}
\cite{tmva} package within the \textsc{\small{ROOT}} \cite{root} framework.
Among all the available discrimination algorithms, this work had initially focused on neural networks 
(\textsl{Multi Layer Perceptron} or MLP) and on \textsl{Boosted Decision Trees} (BDT).
During these early tests, it became immediately clear that the BDT technique 
was more stable and gave better results in terms of signal-to-background discrimination, thus confirming that 
this technique is the best out-of-the-box 
multivariate procedure, requiring minor adjustments to be applicable.

A summary of the operation of the BDT technique is given in Appendix \ref{BDTDesign}, with details
particularly concerning the design of the decision trees, 
the difficulties which can be encountered through overtraining, 
and the best pruning strategies to apply for the case of IACTs.
The multivariate classification technique consists of four main phases: 
the first one is the \textsl{training} (see Appendices \ \ref{TrainingPhase}, \ref{Boosting}) 
where the ``forests" of decision trees are constructed, 
with a set of events randomly chosen from the input.
Then, in a second phase, the classifier performance is \textsl{tested} with an independent set of events, 
and \textsl{evaluated} (see Appendices \ref{Pruning}, \ref{Classifier}).
This study led to a specific design of the decision trees, differing somewhat from the default options 
of the BDT procedure as implemented in \textsc{\small{TMVA}}. 
Since these details could be useful for the implementation of this technique for other IACTs,
they are included in Appendix \ref{TreeDesign}.
After the definition of the final cut values,
the classifier is finally \textsl{applied} to the data (see Appendix \ref{CutAppli}).

\subsection{Simulated \gray\ and measured hadron background data samples} 

The \gray\ shower samples were produced by MC 
simulations with {\textsc{Kaskade}}\footnote{One of the two shower simulation packages used in \hess, 
the other being Corsika\cite{Corsika}.}, 
\cite{Kaskade,GuyPHD,GuyNote}. 
The \gray\ showers are simulated between $30\;\rm GeV$ and $120\;\rm TeV$
for two different azimuths (\textsl{North} and \textsl{South}), 
several zenith angle bands spaced regularly in $\cos(\rm \theta_{zen})$  
(in steps of 0.05) from $0^{\circ}$ to $70^{\circ}$, and 
for a point-like source.
The resulting Cherenkov light is then propagated through the \hess\ detector with 
{\textsc{Smash}}\footnote{This program is one of the two available \hess\ 
detector-response simulation programs.}, \cite{GuyPHD}, 
where the source position can be shifted to different offsets $D$ from the centre of the camera
(six $D$ values spaced by $0.5^{\circ}$ from $0^{\circ}$ 
to $2.5^{\circ}$ are simulated, where $2.5^{\circ}$ corresponds to the radius of the \hess\ field-of-view or FoV).
Taking advantage of the large set of available \hess\ data as a background sample, 
the use of time-consuming and marginally-reliable hadron simulations could be avoided.
The background was therefore extracted from extragalactic observations using 
regions of the camera which do not contain any known \gray\ source,
at various zenith angles, and taking into account the northern or southern
azimuth of the observations (to account for different geomagnetic field effects).
All the hadrons reconstructed out to angles including the \hess\ FoV plus $0.2^\circ$
are taken into account.
\begin{figure}[t]
  {\includegraphics[width=0.45\textwidth]{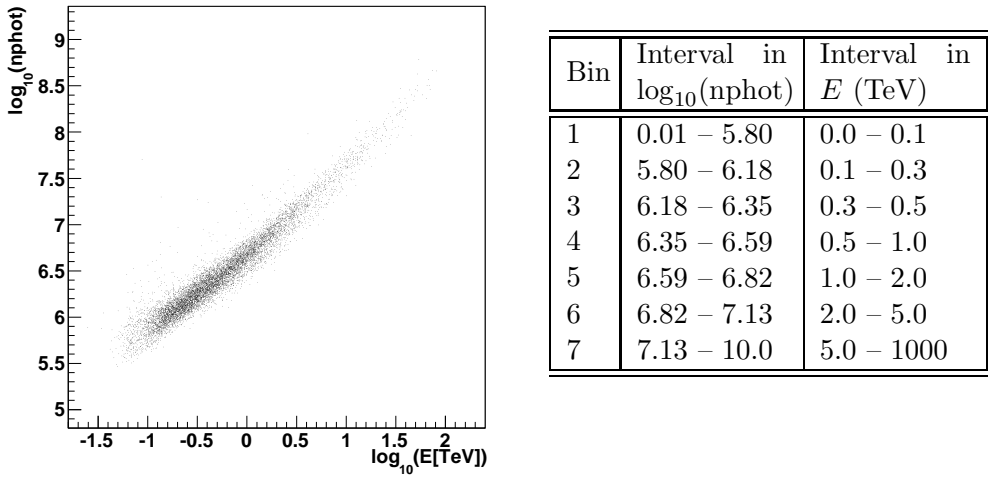}}
  {\begin{minipage}[t]{0.5\textwidth}
      \begin{small}
        \vspace{-6cm}
        \begin{tabular}{|m{0.5cm}|m{2cm}|m{2cm}|}
          \hline
          \hline
          Bin & Interval in $\log_{10}$(nphot) & Interval in $E$ (TeV) \\
          \hline
          \hline
          1 & 0.01 -- 5.80  & 0.0 --  0.1     \\
          2 & 5.80 -- 6.18  & 0.1 --  0.3     \\
          3 & 6.18 -- 6.35  & 0.3 --  0.5     \\
          4 & 6.35 -- 6.59  & 0.5 --  1.0     \\
          5 & 6.59 -- 6.82  & 1.0 --  2.0     \\
          6 & 6.82 -- 7.13  & 2.0 --  5.0     \\
          7 & 7.13 -- 10.0  & 5.0 --  1000    \\
          \hline
          \hline
        \end{tabular}
      \end{small}
      \end{minipage}}
  \caption{\small{\textsl{Left panel:} number of fitted photons in the reconstructed shower 
      (\textsl{nphot}) 
      as a function of the reconstructed energy in TeV\@ (log scales). 
      \textsl{Right panel:} the definition of the \textsl{nphot} bins 
      and the approximate corresponding values (based on the left-hand figure) 
      of the reconstructed energy (in TeV) are illustrated.
  }}
  \label{nphotbins}
\end{figure}

\subsection{Training and Test Configurations} 
\label{TrainingTest}
A set of bins in energy, zenith angle, and telescope multiplicity 
(i.e., the number of triggering telescopes; two to four in the case of \hess) are defined for 
the training/test phases of each of the implemented analysis configurations (see Sec.\ \ref{HardSoftSpectra}). 

There are eight zenith angle bins and seven energy bins, in order to regroup the different classes of events showing
similar properties. The shape of the parameter distributions for each defined group may differ from the other groups,  
leading to a different background rejection performance. 
The zenith angle bins are defined in $10^\circ$ intervals from $0^\circ$ up to $40^\circ$, 
then in $5^\circ$ intervals, and finally in a single interval\footnote{\samepage Note that the definition 
of zenith angle bins which are regularly 
spaced as a function of cos($\theta_{\rm zen}$) as used in the MC simulations is more convenient; 
however, the different choice used here for the \mva\ training has been imposed by 
some initial compatibility constraints with other modules inside the \hess\ Analysis Package.} 
from $55^\circ$ to $90^\circ$
for each of which seven energy bins are defined by the number of photons 
in the shower as fitted by the 3D-model (referred as to \textsl{nphot}), 
as shown in Fig.\ \ref{nphotbins}.
The latter choice has been found to be effective in avoiding brusque variations in the final effective area.

The phase space is additionally subdivided in two bins in telescope multiplicity: 
one for the two-telescope events and one for the three and four-telescope events,
leading to a total of 112 bins altogether. 
This split has been motivated by the fact that for events which trigger only two telescopes, 
less information is available for geometric reconstruction, therefore 
leading to a parameter space which is considerably different with respect to
that for events triggering three and four-telescopes. 
It should be noted that in order to avoid badly-reconstructed \gray\ events polluting 
the training and test phases, such events are excluded using 
a cut on the angular distance on the sky between the 
reconstructed event direction with respect to the source direction 
($< 0.11^{\circ}$).

\begin{figure}[t]
  {\subfloat[] {\includegraphics[width=3in]{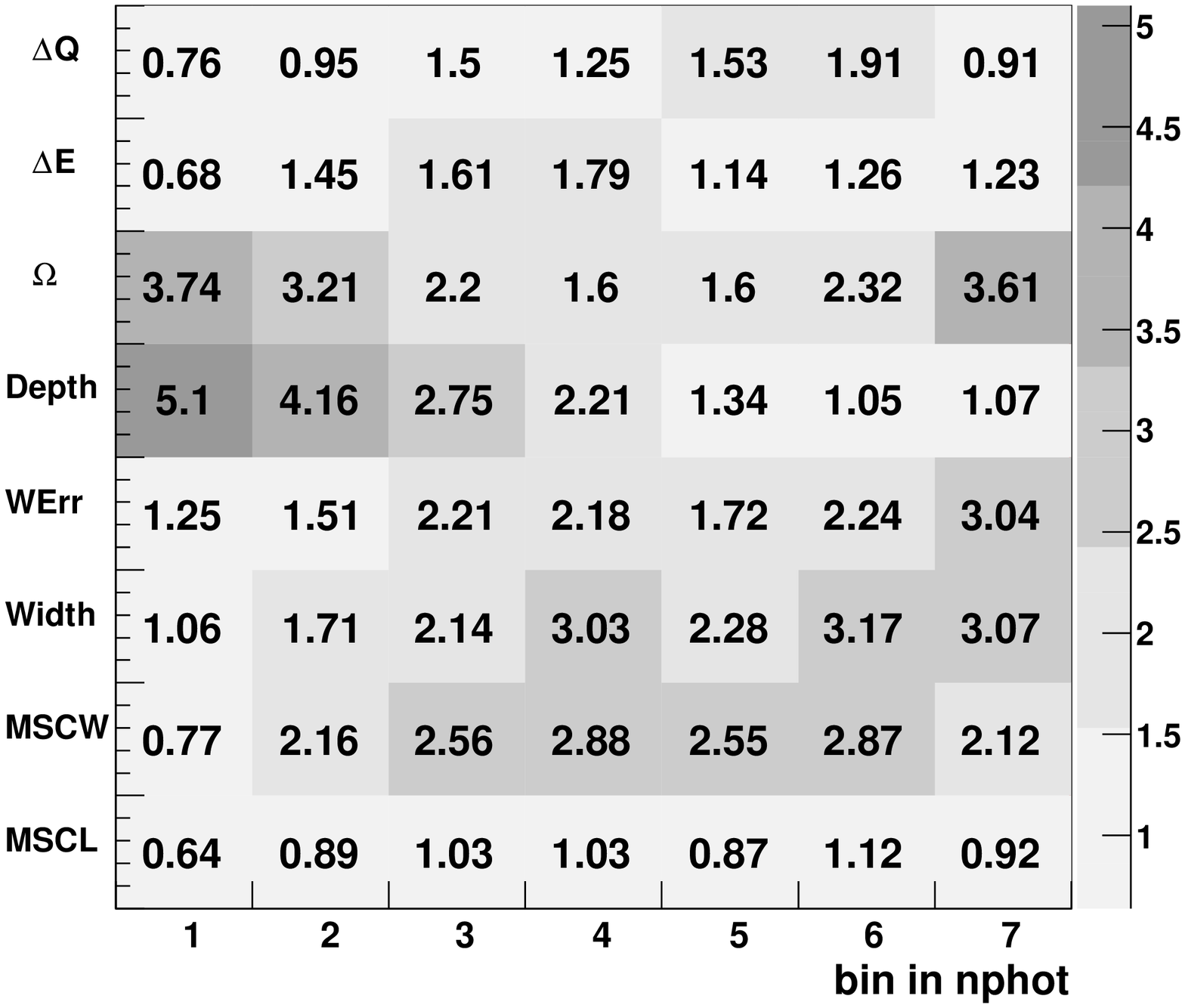}}}
  {\subfloat[] {\includegraphics[width=3in]{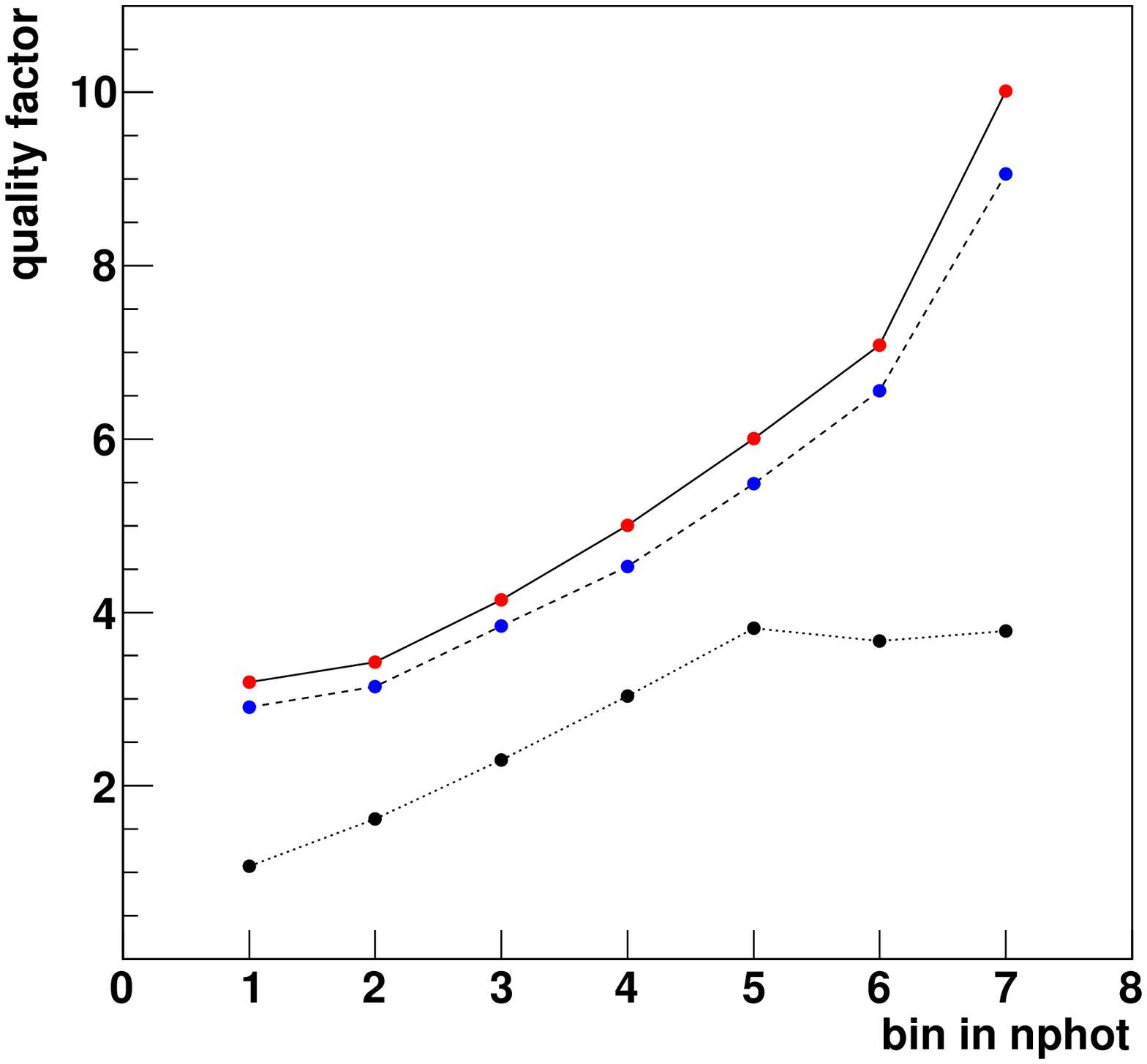}}}
  \caption{\small{(a) Two-dimensional plot of the importance of the parameters 
      during training as a function of the bin in \textsl{nphot}. 
      The value shown in each box is the sum over all zenith angle bins 
      of the relative weights, 
      given by the BDT training to each discriminant parameter (see text).
      \textsl{Depth}, \textsl{ErrW} and \textsl{RedW} refer to the 3D-depth, the 3D-width-err 
      and the 3D-width, respectively 
      (b) Quality factor (QF) as a function of the \textsl{nphot} bin 
      for given \gray\ efficiencies, at a zenith angle of $0^\circ$ 
      and for a configuration having a charge threshold of $80\;\rm p.e.$ per telescope.
      Dotted, dashed and continuous lines represent the QF obtained with the following configurations of parameters:
      a first group (MSCL, MSCW), 
      a second group (adding three 3D-model parameters), 
      and a third group (adding the three new newly defined parameters), 
      respectively. 
  }}
  \label{VariableImportance}
\end{figure}

\begin{figure}[t]
  {\subfloat[] {\includegraphics[width=3in]{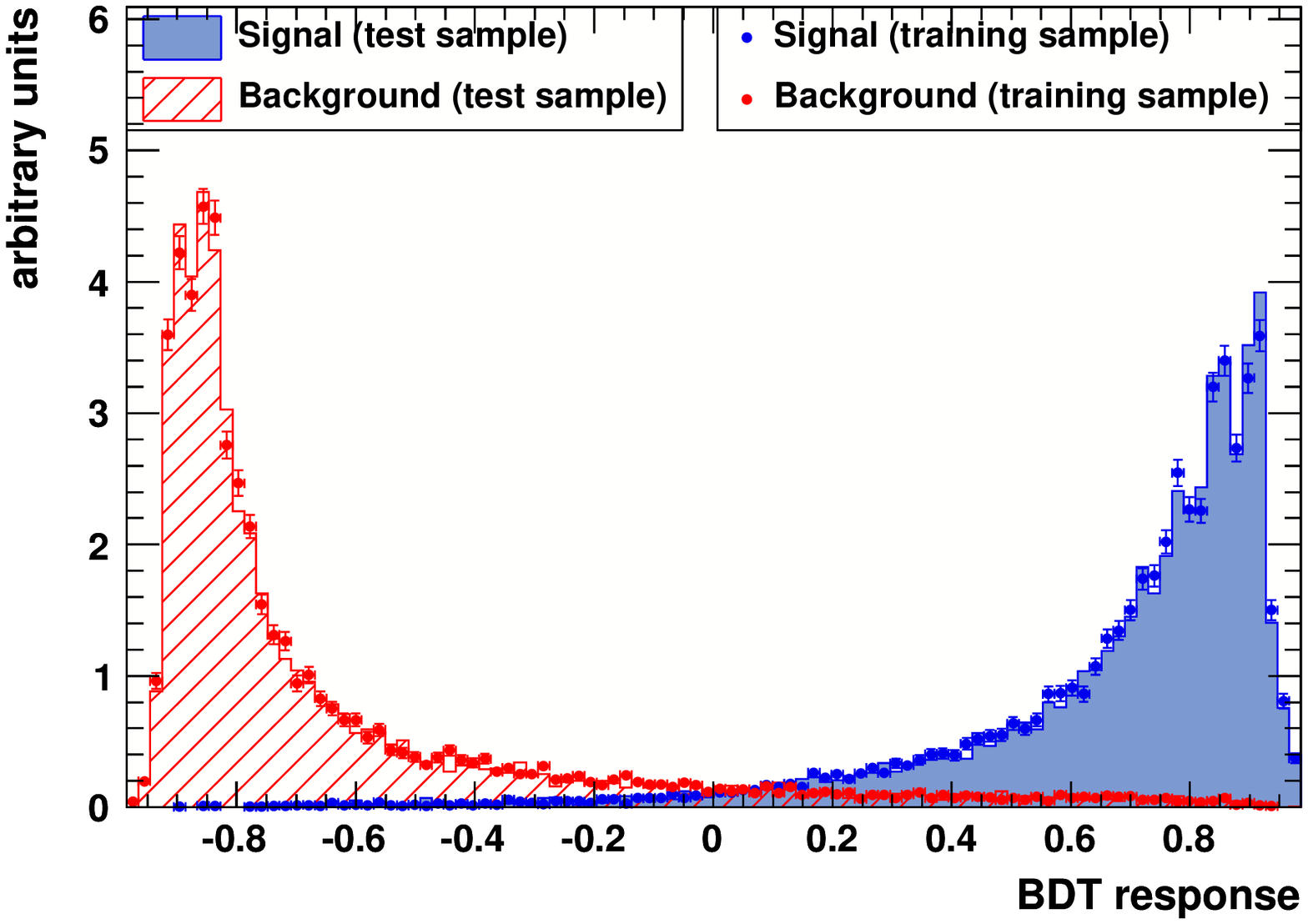}}}
  {\subfloat[] {\includegraphics[width=3in]{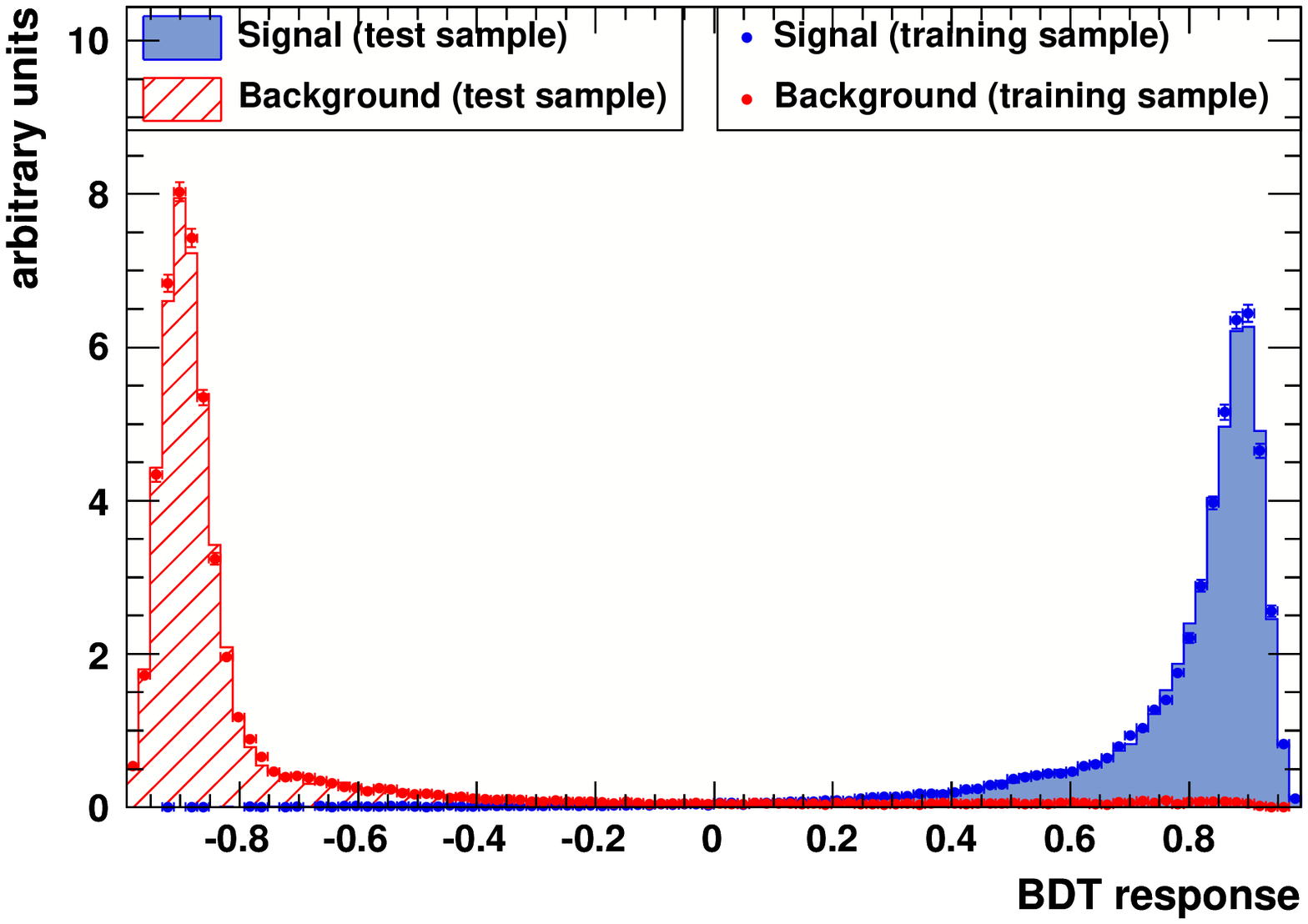}}}
  \caption{\small{
      The BDT classifier 
      response to the BDT training and test phases 
      for the bin between $10$--$20^\circ$ in zenith, 
      and between 1--$2\;\rm TeV$ for two-telescope events (a) and for three-telescope 
      and four-telescope events (b). 
      The response to both the training and test samples are shown (points ans histograms respectively).
  }}
  \label{BDTresponse}
\end{figure}

\subsection{Relative importance of the parameters for discrimination} 

\label{importance}
The multivariate method \textsl{ranks} the discriminant parameters 
according to the frequency with which they are used in the splitting
of the decision tree nodes. 
The values shown in the boxes of Fig.\ \ref{VariableImportance}a
are given by the sum of the relative weights given by the BDT training.
It can be observed that at low energies ($E < 500\;\rm GeV$) the most important 
discriminant parameter is 
the 3D-model estimated depth of the shower maximum, with its importance decreasing 
as a function of energy.
The second parameter in order of importance is $\ensuremath{\Omega}$: 
while its discrimination performance is clearly high in all energy bins, 
it is more powerful at low energies\footnote{Though $\ensuremath{\Omega}$ is still less  
discriminant than the depth at shower maximum.} ($E < 500\;\rm GeV$, bin in \textsl{nphot} $\leq3$)
or high energies ($E > 5\;\rm TeV$, bin in \textsl{nphot} $\geq7$).
The least discriminant parameter in almost all cases is MSCL.
The importance of the chosen parameters can also be evaluated by checking the relative quality factor 
when progressively adding groups of variables to the training, where the quality factor is defined as 
\begin{equation}
Q = \frac{\epsilon_{\gamma}}{\sqrt{\epsilon_{\rm bkg}}}, \; {\rm where} \; \epsilon_{\gamma,\rm bkg} =  N_{\rm cut}/N_{\rm reco}   
\end{equation}
with $N_{\rm{cut}}$, $N_{\rm{reco}}$ being the number of events after and before cuts, respectively, in a given bin. 
Three different test configurations are defined: the first group includes only MSCL and MSCW; 
the second comprises MSCL, MSCW plus the group of the three chosen 3D-model parameters; 
and the third adds the three new ones defined for the purpose of this work to the preceding groups.
The results are given in Fig.\ \ref{VariableImportance}b: 
a significant gain in performance is achieved when adding the group of the three 3D-model
parameters to the Hillas-parameter based ones; and 
with the addition of the new parameter group a further $\sim 20\%$ of background is cut away, 
leading to an additional gain in $Q$ of $\sim 12\%$. 

Finally, it should be noted that investigation of the effects of the 
correlations between the discriminant parameters resulted in 
the elimination of a few parameters which had been defined 
at an early stage of the analysis development (see \cite{YBICRC}).
Optimization tests within the specific framework used here showed 
that the inclusion of pairs of highly-correlated variables resulted in a loss of discrimination performance,
probably due to the overall development of the trees in the forest 
being more sensitive to fluctuations.\footnote{Tests of the \textsc{\small{TMVA}} functionality which 
implements parameter decorrelation provided inconclusive results, 
though this may be explored in further developments.}
\begin{figure}[t]
 \centering{
   {\subfloat[] {\includegraphics[width=3in]{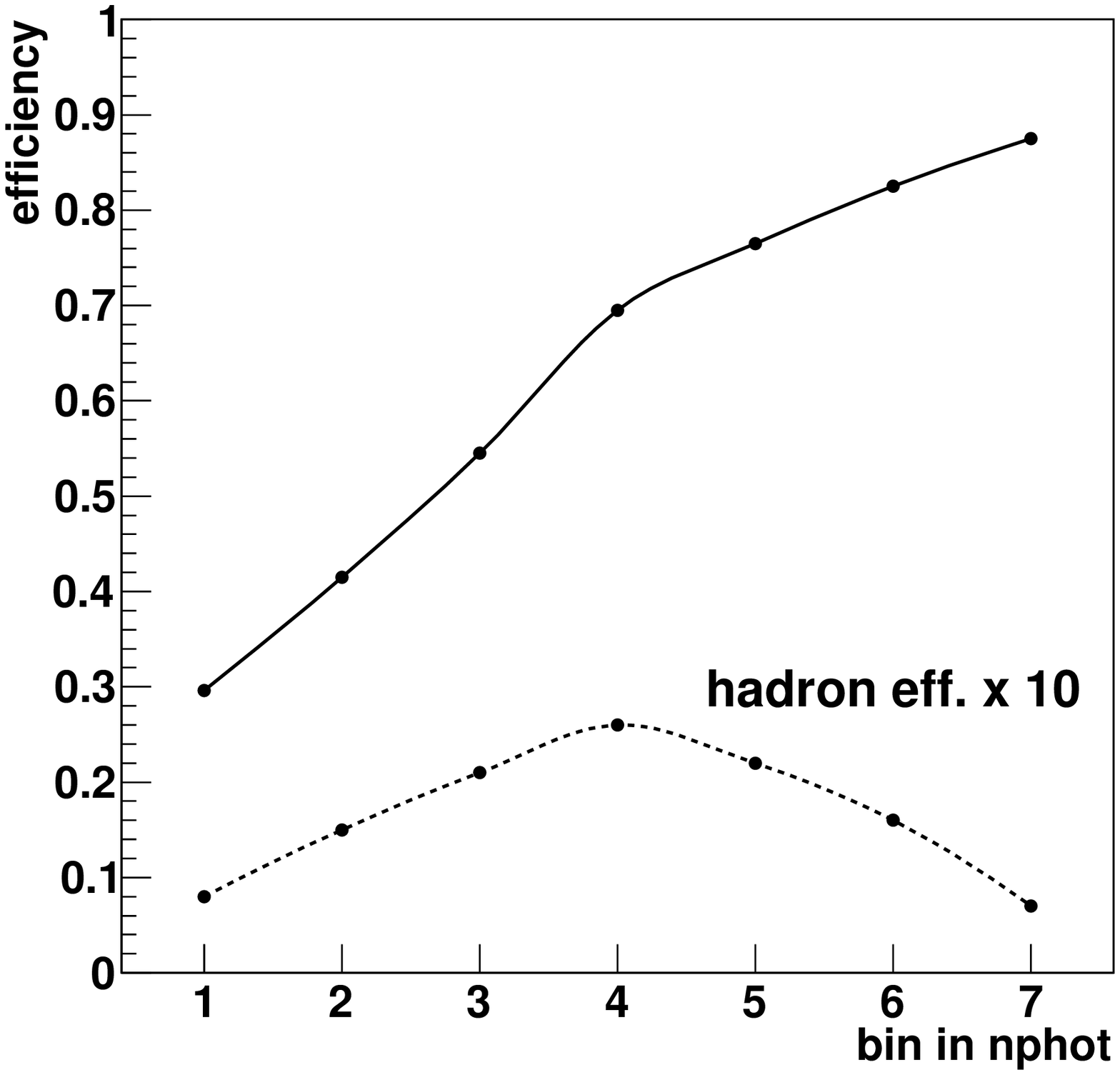}}}
   {\subfloat[] {\includegraphics[width=3in]{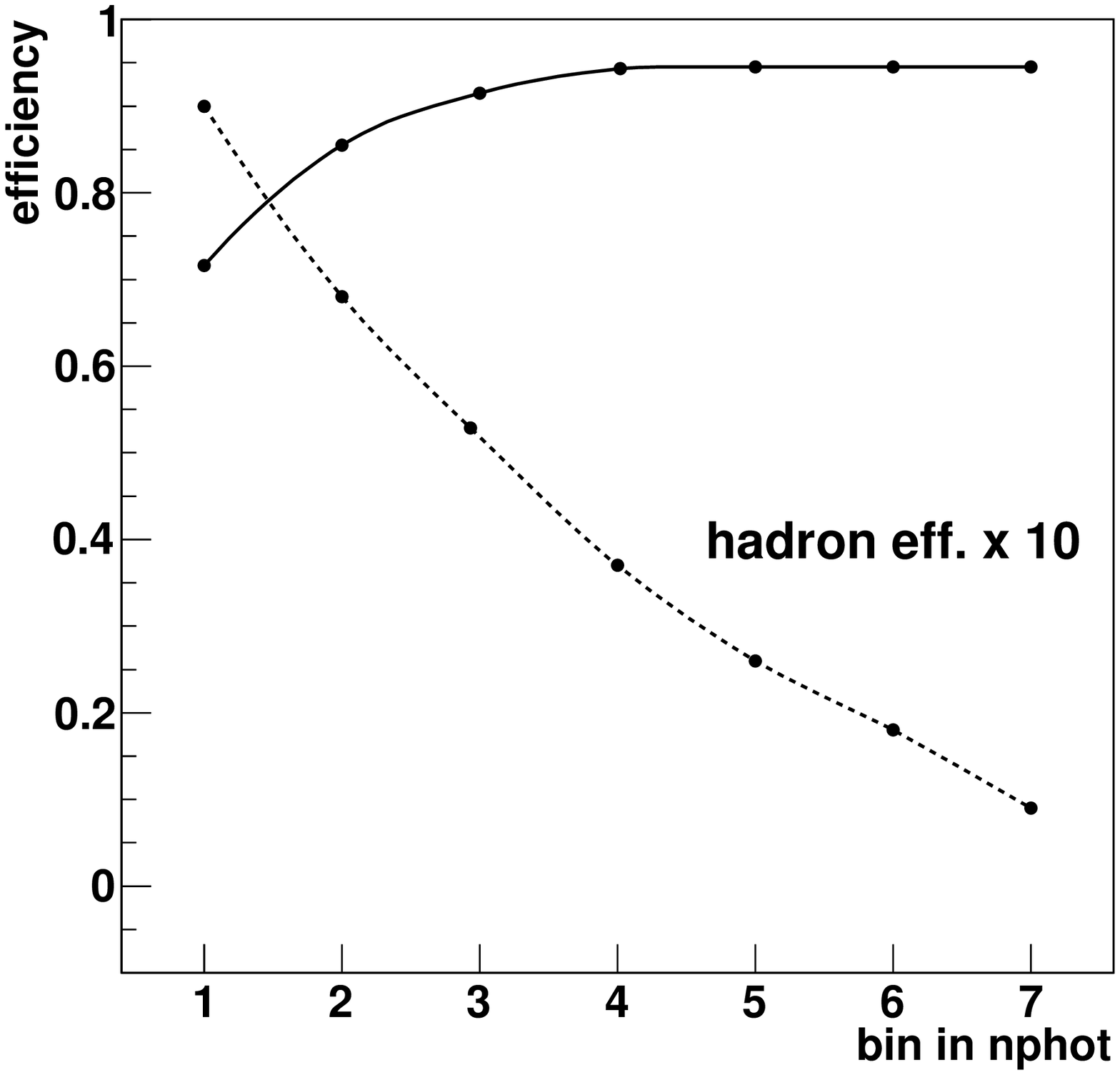}}}}
      \caption{\small{
          Discrimination performance for the analysis configuration having a 110 p.e.\ charge threshold. 
          The solid lines show the \gray\ efficiencies within the angular pre-cut
          mentioned in Par.\ \ref{TrainingTest},
          while the dashed lines show the corresponding hadron efficiencies multiplied by 10. 
          These are shown as a function of the \textsl{nphot} bin for a zenith angle of $0^\circ$,
          (a) for two-telescope events and (b) for three and four-telescope events. 
          The definition of the photon bins (\textsl{nphot}) can be found in Fig.\ \ref{nphotbins}.}}
      \label{HardSpectra}
\end{figure}

\subsection{Definition of cuts for different source types} 

\label{HardSoftSpectra}

The training on each event class described in Sec.\ \ref{TrainingTest} 
provides a separate BDT classifier (see Fig.\ \ref{BDTresponse} for two examples), 
on which the analysis cut is based.
For the definition of the final cut values, the use of the highest significance 
criterion\footnote{Which depends on the relative normalization  
between signal and background.}
provided by the BDT training process
was found to be subject to statistical fluctuations.
For this reason, \gray\ efficiency profiles as a function of energy 
were defined in order to optimize the detection sensitivity for sources with different fluxes,
through an iterative optimization process.
For illustration, the pre-defined \gray\ and corresponding background efficiencies
for one analysis configuration, of the five described below, is shown in Fig.\ \ref{HardSpectra}. 
For the purpose of the analysis of sources showing different spectral indices,
five configurations are defined based on the following charge thresholds: 
40, 60, 80, 110 and $150 \;\rm p.e$\@. 

To achieve an optimal sensitivity for sources with spectral indices\footnote{Given a power-law photon 
differential flux in $E_{\rm TeV}^{-\Gamma}$.} $\Gamma \lesssim 3$ 
and fluxes above $\simeq 1\%$ C.U.,
an analysis configuration
having a charge threshold of $80\;\rm p.e.$ (and a similar one at $60\;\rm p.e.$) has been developed. 
This configuration is therefore adapted for most Galactic-plane observations, 
where most of the sources discovered have hard spectra (see \cite{GPS})
and is also used for low-redshift extragalactic observations. 
For similar spectral indices but lower fluxes ($< 1\%$ C.U.), another configuration has been defined with
a charge threshold of $110 \;\rm p.e.$ and for which 
the \gray\ efficiencies have been set so as to yield a lower rate of low-energy two-telescope events 
than the other configurations, see Fig.\ \ref{HardSpectra}.
These two characteristics help in reducing the lower 
energy fluctuations due to the background events, thus 
enhancing the sensitivity for this source class.
For morphological studies of extended sources, 
a dedicated configuration has been developed which has an enhanced PSF, 
optimizing the performance in geometrical reconstruction by raising 
the charge threshold to $150\;\rm p.e.$

For high-redshift extragalactic observations where the effect 
of \gray\ absorption becomes significant, steep-spectrum ($\Gamma \sim 3$) sources are naturally expected.
To optimize the sensitivity for the detection of this class of sources, the image size threshold has been 
lowered to $40$ p.e. in order to enhance the signal of the abundant lower energy events.

App.\ \ref{PSFcuts} discusses 
the instrument response functions associated with these configurations.

\begin{figure} 
  {\subfloat[]{\includegraphics[width=3in]{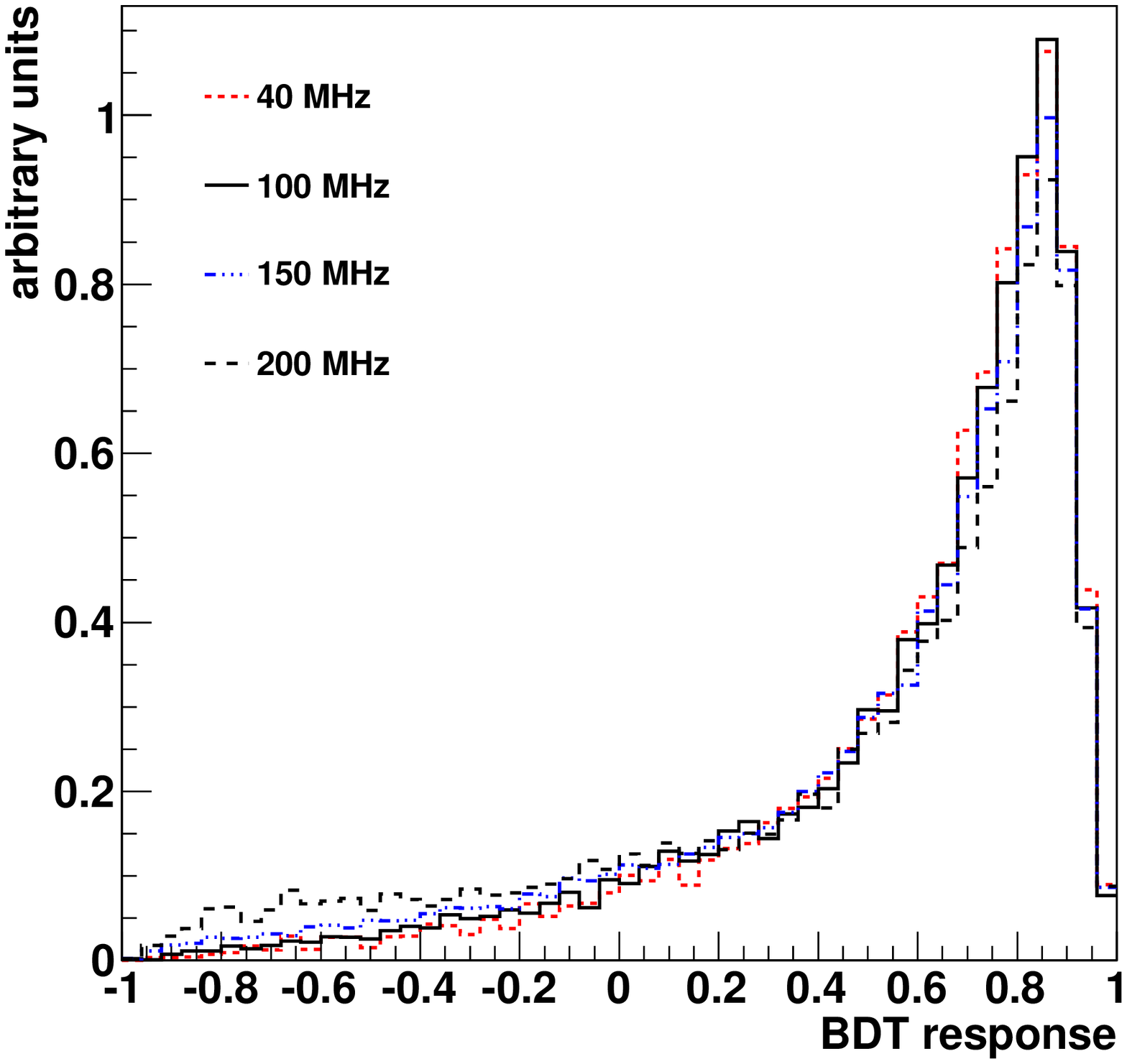}}}
  {\subfloat[]{\includegraphics[width=3in]{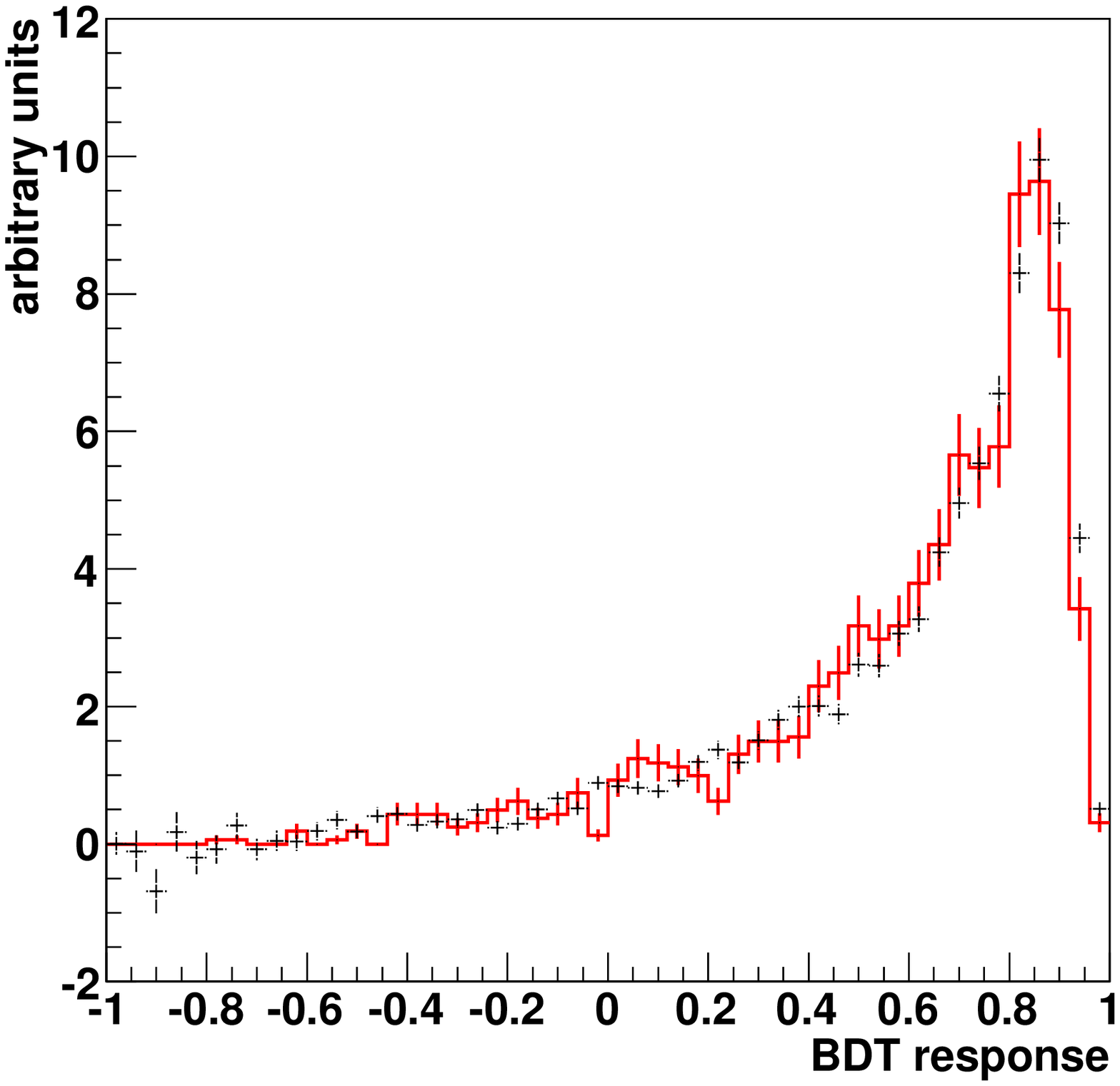}}}
  \caption{\small{
      (a) BDT response for simulated \grays\ for different NSB rates ranging from 40 to {200\;\rm MHz}. 
      The sample was simulated at Zenith and follows a power-law with differential index $\Gamma=2$.  
      The BDT response remains stable for NSB rates up to {150\;\rm MHz},  
      the relative migration of events from the signal to the background range remaining well below 2\%. 
      For a larger rate of {200\;\rm MHz}, the latter attains $\sim$10\%, 
      without significant impact on the measurements (see text). 
      (b) BDT response for \gray\ simulated events (solid line) 
      and real \gray\ events (crosses with error bars) from the Crab nebula.
    }}
  \label{BDTNSB}
\end{figure}

\begin{figure} 
  \centering{\includegraphics[width=4.6in]{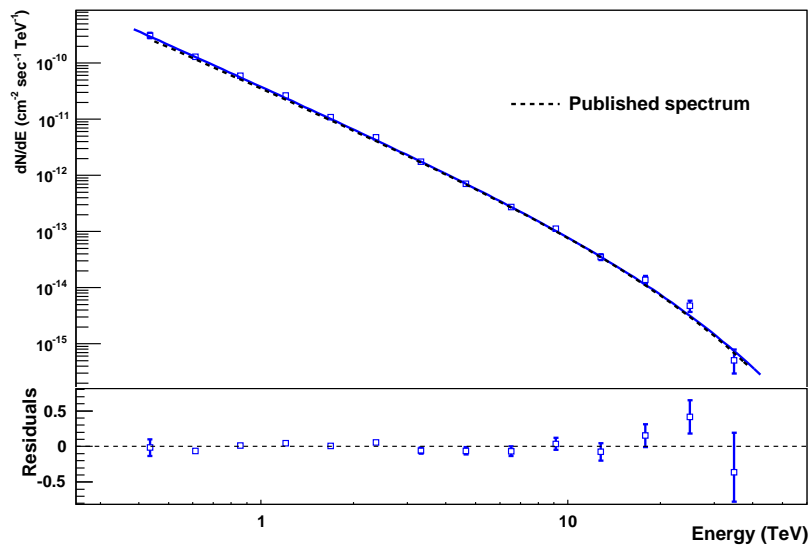}}
  \caption{\small{
      Reconstructed spectrum of the Crab nebula (points), showing an excellent 
      agreement within the systematic errors 
      with the published results (dashed line)\cite{crab}. 
      The residuals are calculated with respect to the fitted spectral shape.
  }}
  \label{BDTperf}
\end{figure}

\section{Robustness and consistency checks}
\label{Robu}

\subsection{Robustness with respect to variations of the NSB rate}

One of the key points of the present approach is its ease of implementation
without the need for calibration of the discriminant variables as a function of the NSB rate 
for a given FoV.  
With the \hess\ telescopes, the NSB median value of the latter can vary from few tens of 
MHz
per pixel for extragalactic observations, up to $\sim{130\;\rm MHz}$ 
for $95\%$ of \hess\ observations
and for central parts of the Galactic plane. 
Fig.\ \ref{BDTNSB}a shows the distribution of the BDT response 
for simulated \grays\ following a power-law ($\Gamma=2$) at Zenith, 
with different NSB rates ranging from 40 to {200\;\rm MHz}. 
The stability of the BDT response to NSB rates up to {150\;\rm MHz} is remarkable: 
the relative migration of events from the signal to the background range remains well below 2\%. 
For a larger NSB rate of {200\;\rm MHz}, the migration attains $\sim$10\%, 
however, the resulting systematic error on the effective area (i.e., $10\%$) 
remains dominated by the error 
due to the uncertainty in the energy scale \cite{crab},  
depending on the spectral index of the source.
A check of the energy dependence of such an effect has also been carried out
through reconstruction of simulated \grays\ following the same power-law 
and the same NSB levels as above.
No significant bias was found on the measured spectral index. 
This was expected, since the choice of those variables already available 
(MSCL, MSCW, 3D-width, 3D-width-err, 3D-depth, see Sec.\ \ref{HillasPara} and \ref{ModelPara}), 
and the definition of the new discriminant parameters 
($\Omega$, $R_E$ and $\Delta Q$, see section \ref{NewPara}) 
was made following the study of their individual robustness with respect to 
the NSB rate. 

\subsection{Consistency checks using real \gray\ data}

The consistency of the BDT response was checked by comparison of its distribution 
for simulated  \grays\ to that of the signal from the Crab nebula. 
MC simulations were made for 
\grays\ with a spectrum following a power-law up to {100\;\rm TeV} with a differential photon index $\Gamma=2.6$,  
and a zenith angle of $46^\circ$, matching closely 
the chosen data-set.\footnote{The NSB simulated rate is {100\;\rm MHz} per pixel, 
close to the median value measured on the Crab nebula FoV as seen by \hess}
As can be seen on Fig.\ \ref{BDTNSB}b, there is a good agreement between the simulated 
sample's BDT distribution and that of the Crab nebula signal.

Finally, as is mandatory in VHE \gray\ astronomy, an overall consistency check 
of the analysis chain was performed by comparing the spectrum 
as measured with the present \mva\ approach to that obtained 
with the standard Hillas-parameter based one, for a source such as the Crab nebula \cite{crab}. 
As shown in Fig.\ \ref{BDTperf}, fitting an exponentially cut-off power-law to the data in the 
range {380\;\rm GeV} to {44\;\rm TeV} using the method in \cite{Piron} 
yields a flux of $I_0 =(4.31 \pm 0.07_{\rm stat}) \times 10^{-11} \mathrm{cm}^{-2} \,
\mathrm{s}^{-1} \, \mathrm{TeV}^{-1}$, an index of $\Gamma=2.43\pm0.03$, and a 
cut-off energy $E_c=15.4\pm 2.2\;\rm TeV$.  
These values are fully compatible with those published, evaluated
in the range {410\;\rm GeV} to {40\;\rm TeV}: 
$I_0 =(3.76 \pm 0.07_{\rm stat}) \times 10^{-11} \mathrm{cm}^{-2} \, \mathrm{s}^{-1} \, \mathrm{TeV}^{-1}$,
$\Gamma=2.39\pm0.03$, and $E_c=14.3\pm 2.1\;\rm TeV$, 
noting that the data-set used here is twice as large as that in \cite{crab}
and that the difference ($\sim$15\%) in the differential flux is well within 
the systematic error of the IACT measurements. 
The same type of check made on other established sources 
yielded an excellent level of compatibility, further 
reinforcing the confidence in the approach presented here.

\section{Performance with published \hess\ sources}

\label{Perf}

\begin{table}[t]
\begin{center} 
\begin{footnotesize}
\begin{tabular}{|m{2.3cm}|m{1.2cm}|m{0.7cm}|m{0.6cm}|m{2.2cm}|m{0.7cm}|m{0.6cm}|m{0.6cm}|m{0.6cm}|m{0.5cm}|m{0.7cm}|}
\hline
\hline
Source & $\Gamma$ & $\phi_0$ C.U. & $\theta_{\rm zen}$ & Analysis & LT & On & Off &  N$_{\gamma}$ & N$_{\sigma}$ & $\sigma/\sqrt{\rm h}$ \\
\hline
\hline
\hline
\hline
$\rm PG\;1553+113$ & $4.5$      & 3.4\% & $40.0^\circ$ & Hillas 40 p.e. & 24.8 & 5748 & 4962 & 785 & 10.2 & 2.05 \\ 
\cite{ES1553} & ~$\rm \pm0.3_{stat}$ &  &  & MVA 80 p.e. & 24.8 & 409  & 269  & 140 & 7.6  &  1.51 \\ 
 & ~$\rm \pm0.1_{sys}$  &  &  & MVA 40 p.e. & 24.8 & 1015 & 665  & 349 & 11.9 &  {\bf 2.39}  \\ 
\hline
\hline
$\rm 1ES\;0347-121$  & $\rm 3.10$ & 2.0\% & $19.8^{\circ}$ & Hillas 80 p.e. & 25.4 & 1167 & 840 &  327 & 10.1 & 2.00  \\
\cite{ES0347} & ~$\rm \pm0.23_{stat}$ & & & MVA 80 p.e. & 25.4   &  490    &  263   &  226   & 11.8 & 2.35 \\
 & ~$\rm \pm0.10_{sys}$  & & & MVA 40 p.e. & 25.4   &  1082   &  706   &  375   & 12.4 & {\bf 2.46} \\
\hline
\hline
$\rm H\;2356-309$ & $\rm 3.06$ & 1.6\% & $19.0^{\circ}$ & Hillas 80 p.e. & 116.8 & 8899 & 7718 & 1185 & 12.6 & 1.16  \\
\cite{H2356} & ~$\rm \pm 0.15_{stat}$ & & & MVA 80 p.e. & 116.8 & 2152 & 1319 & 833  & 20.0   & {\bf 1.85} \\
 & ~$\rm \pm 0.10_{sys}$ & & & MVA 40 p.e. & 116.8 & 4477 & 3277 & 1199 & 18.9 & 1.74 \\
\hline
\hline
$\rm Cen \;A$ & $\rm 2.7$ & 0.8\% & $24.5^{\circ}$ & Hillas 80 p.e. & 120.0 & 4199 & 3868 & 330 & 5.0 & 0.46 \\
\cite{CenA}   & $\rm \pm 0.5_{stat}$ & & & MVA 80 p.e.  & 120.0 & 1437 & 1109 & 324 & 9.0 & {\bf 0.82} \\
              & $\rm \pm0.2_{sys}$ & & & MVA 40 p.e.  & 120.0 & 3306 & 2967 & 338 & 5.9 & 0.54 \\
\hline
\hline
$\rm G0.9+0.1$ & $\rm 2.4$ & 2.0\% & $21.5^{\circ}$ & Hillas 200 p.e. & 50.1 & 606 & 310 & 296 & 14.4 & 2.03  \\
\cite{G0.9}    & $\rm \pm0.11_{stat}$ & & & MVA 110 p.e. &  50.1  & 731   &  312   & 419  & 19.3 & {\bf 2.73} \\
               & $\rm \pm0.20_{sys}$ & & & MVA 80 p.e.   &  50.1  & 1010  &  553   & 457  & 16.8 & 2.37 \\
\hline
\end{tabular}
\label{TablePerformance}
\caption{\small{Performance of the analysis on some published \hess\ sources. 
  For each, published results obtained with Hillas-parameter based analyses are shown in the 
  first line and results obtained here with the 
  \mva\ method presented in this paper are shown just below.
  $\Gamma$, $\phi_{0}$ and are the published differential spectral index and flux in C.U., 
  $\theta_{\rm zen}$ is the mean zenith angle of observation in degrees, ``LT'' is the livetime in hours, 
  ``On'' represents the total number of events around the position of the source in the sky, 
  ``Off'' represents the normalised number of background events,
  $N_{\gamma}$ is the number of excess events,
  $N_{\sigma}$ is the significance of the detection and $\sigma/\sqrt{\rm h}$ is the significance 
  over the square root of the observation livetime in hours, for which the best-adapted configuration is shown in bold. 
  The normalisation factor ($\alpha$ in Eqn.\ 17 of \cite{lima}) depends on 
  the source observation conditions (the offset, the angular cut between the reconstructed direction 
  and the source position, 
  and the disposition of excluded regions in the FoV), and varies from 0.06--0.125 for the sample shown.}
}
\end{footnotesize}
\end{center} 
\end{table} 

The performance of the analysis scheme proposed here has been tested on a set of faint \hess\ sources. 
The spectral index is used to select, a priori, 
two test configurations which are expected to be the best adapted. 
The results obtained for a sample of five cases are presented in Tab.\ 1 and compared to the standard analysis 
(labelled as Hillas), so as to illustrate the trends as a function of the sources' spectral indices.

There are four extragalactic sources ordered by spectral index --
the three blazars $\rm PG\;1553+113$, $\rm 1ES\;0347-121$ and $\rm H\;2356-309$, 
and the radio-galaxy $\rm Cen\;A$ 
-- together with one hard-spectrum Galactic source, the composite 
supernova remnant $\rm G0.9+0.1$ (references are given in the table). 

The main trend which can be seen is that the best-adapted \mva\ configuration 
for the source with the softest spectrum ($\rm PG\;1553+113$) 
is the one with the lowest charge threshold (i.e., $40\;\rm p.e.$), 
whereas the $110 \; \rm p.e.$ configuration yields the best performance 
for the hardest spectrum source ($\rm G0.9+0.1$).
The $80\;\rm p.e.$ and $40\;\rm p.e.$ configurations 
are somewhat equivalent for sources with spectral indices $\Gamma\sim 3$, 
a value which appears to constitute a pivot index above which the use of the lowest 
charge threshold configuration is the best adapted.

In more detail, indeed, the \mva\ $40\;\rm p.e.$ and $80\;\rm p.e.$ 
configurations give a similar gain in sensitivity over the standard Hillas-parameter based analysis
for the two blazars for which $\Gamma\sim 3$
(by a factor of about $1.2$ for both configurations for $\rm 1ES\;0347-121$,
and 1.5--$1.6$  for $\rm H\;2356-309$). 
In contrast, for $\rm PG\;1553+113$, \mva\ at $80\; \rm p.e.$ 
is significantly less efficient than that at $40 \; \rm p.e.$\footnote{For the very soft spectrum 
source $\rm PG\;1553+113$, the reference in the table also applies specific Hillas \textsl{soft spectrum} cuts with
a $40 \; \rm p.e.$ threshold, similarly to the best \mva\ configuration here.}
On the other hand, for $\rm Cen\;A$, which shows a harder spectrum ($\Gamma \simeq 2.7$) than the other extragalactic sources, 
the \mva\ $80\;\rm p.e.$ configuration becomes dominant over the $40\;\rm p.e.$ one, 
with a clear gain in sensitivity as compared to the standard Hillas analysis 
by a factor of $1.8$ (versus a factor $1.2$ for the $40\;\rm p.e.$ case).

With a gain in sensitivity of the \mva\ approach over the Hillas-parameter based analysis 
ranging from a factor $1.2$ to $1.8$ for the sample of sources shown here,
this corresponds in a gain in observation time by a factor $1.4$ to $3.2$, 
as can be seen from the column $\sigma/\sqrt{\rm h}$ of Tab.\ 1.

\section{Conclusions} 

A new analysis method is presented,
based on an \mva\ approach optimized for the detection of low-flux sources, 
conceived so as to be easily and quickly adaptable to other current and future IACTs.

Three new discriminant variables are combined with 
the already-known Hillas-param\-eter based variables mean-scaled image width and length (MSCL, MSCW), 
and with carefully-selected parameters from the more recent 3D-model 
(the 3D-width of the Cherenkov photosphere, its error, and the 3D-depth of shower maximum) 
in a multi-variate discrimination procedure using \textsl{Boosted Decision Trees}, 
where this analysis has been adapted to the particularities of detection with IACTs.
Two of the new discriminant parameters -- one of which uses a new energy reconstruction method described here -- 
are based on the Hillas moments calculated from the 3D-model predicted images in the different telescopes, 
which provides a means to take into account the inter-telescope correlations
which are missing from a simple Hillas-parameter based discrimination framework.  
For the third new discriminant parameter, the energy evaluation method is inverted to predict the 
expected image charges and compare them in a least squares with the images measured in each telescope.

A number of BDT classifiers are defined, in order to take account of the different classes of data,
with bins in zenith angle, telescope-multiplicity and reconstructed energy. 
Note that the number of photons in the Cherenkov photosphere as fitted by 
the 3D-model is used as a proxy for the reconstructed energy as this is found to be effective in avoiding
brusque variations of the final effective area.
The robustness of the new analysis method is shown with respect to the variations 
in NSB for MC simulated \grays\ by examining the evolution of the BDT response. 
The consistency is checked by comparing the BDT response with that 
for the real \gray\ excess from the Crab nebula as seen by \hessend
A final consistency check is performed by comparing the spectrum from the Crab nebula as derived by the present \mva\
approach to that published by \hess\ using the Hillas-parameter based analysis, 
and the two spectra are shown to have an excellent compatibility.
The performance of the different groups of parameters when applied to MC simulated \grays\ and real background data 
is presented, showing the relative gains achieved in each case above the standard Hillas-parameter based analysis.
For the application to the analysis of real sources, several configurations have been defined 
so as to allow cut optimization for different expected source characteristics 
(intensity and spectral hardness), as well as for studies of source morphology.
The method is applied to a set of known \hess\ sources with measured characteristics.
The gain in sensitivity ($1.2$ to $1.8$) is in conformance 
with the expectation based on their spectra and intensities. 

The method is currently being applied successfully
to \hess\ observations, contributing to discovery of new faint sources, see e.g., \cite{TexasAGN}.
The overall development of this new \mva\ approach presented here
has been carried out in order to be easily adapted and applied
to other current and future stereoscopic arrays -- such as \veritas,
and in the near future the \hesstwo\ upgrade (in the stereo energy regime)
-- this ease of application being achieved thanks to the rapidity and flexibility
of the method presented.  These features have been recently demonstrated 
through successful application for the future Cherenkov Telescope Array (CTA) project.

Future improvements of the method include a deeper study of the parameter decorrelation
prior to the use of the \mva\ training procedure, as mentioned in Sec.\ \ref{importance} and
a further investigation on the possibility to mix the two reconstructed directions
(the Hillas-parameter based and the 3D-model ones) as a function of an observable related to the energy
(see Sec. \ref{Stereo}).
The study of the stability and performance of the method as a function
of a degraded performance of the telescopes through MC simulations 
constitutes also an interesting point to be investigated in a future work.

\begin{appendices}

\section{Decision tree generalities and design} 
\label{BDTDesign}

\subsection{Training phase} 
\label{TrainingPhase}

The \textsl{training} phase of a multivariate background rejection method consists of the construction
of a decision tree (see Fig.\ \ref{BDTree}a) given a pre-defined specific binary tree architecture.
The separation of the events into signal and background is carried out starting from a root node 
using the parameter giving the best separation performance and
according to a chosen splitting criterion.
The decision tree continues to grow until its pre-defined maximum size has been reached 
or a minimum number of events are present in the node.

\subsection{Boosting} 
\label{Boosting}

A single decision tree is, however, unstable with respect to statistical fluctuations, so, 
once the first tree is grown, 
the same procedure is repeated in order to build an ensemble 
of decision trees (called a \textsl{forest}) by reweighting the same sample of events 
(this step is called the BDT \textsl{boosting}). 
The \textsl{boosting} has the important feature of
contributing greatly to the statistical stability of the algorithm and 
further improving the separation performance.

\subsection{Overtraining and pruning} 
\label{Pruning}

A key point during the \textsl{training} phase is the need to avoid the so-called \textsl{overtraining} 
of the multivariate procedure: 
the \textsl{overtraining} phenomenon occurs when the separation is too powerful for the training sample,
generally resulting in the overestimate of the real discrimination power.
If \textsl{overtraining} occurs, the multivariate method is then not 
capable of learning the real trend in the data, but simply memorizes the data by heart 
resulting in a poor generalization of the problem. 
When in presence of \textsl{overtraining}, which leads to a too-good fit to the training data,
the overall performance of the method on an independent set of data is generally bad.
So, in order to maximize the separation performance algorithm 
and also to avoid keeping statistically insignificant nodes during the decision process, 
a technique known as \textsl{pruning} is applied to the tree by
cutting back the poorly populated nodes from the bottom part where the tree has reached its
maximum size, up to the root node.

\subsection{Classifier output} 
\label{Classifier}
\noindent 
At a final stage of the BDT procedure, 
all the trees are combined together resulting in a single classifier output given 
by the average (or weighted) behaviour of the individual decision trees.
The classifier resulting from the \textsl{training} phase is tested to evaluate 
the global performance of the multivariate method and this is done during the \textsl{test} phase.
During the \textsl{testing}, an independent set of events is used as input for the splitting procedure so that
the \textsl{training} classifier distribution can be compared with that resulting from the \textsl{test}.
\begin{figure}[t]
  {\hspace{-0.9cm}\includegraphics[width=0.6\textwidth]{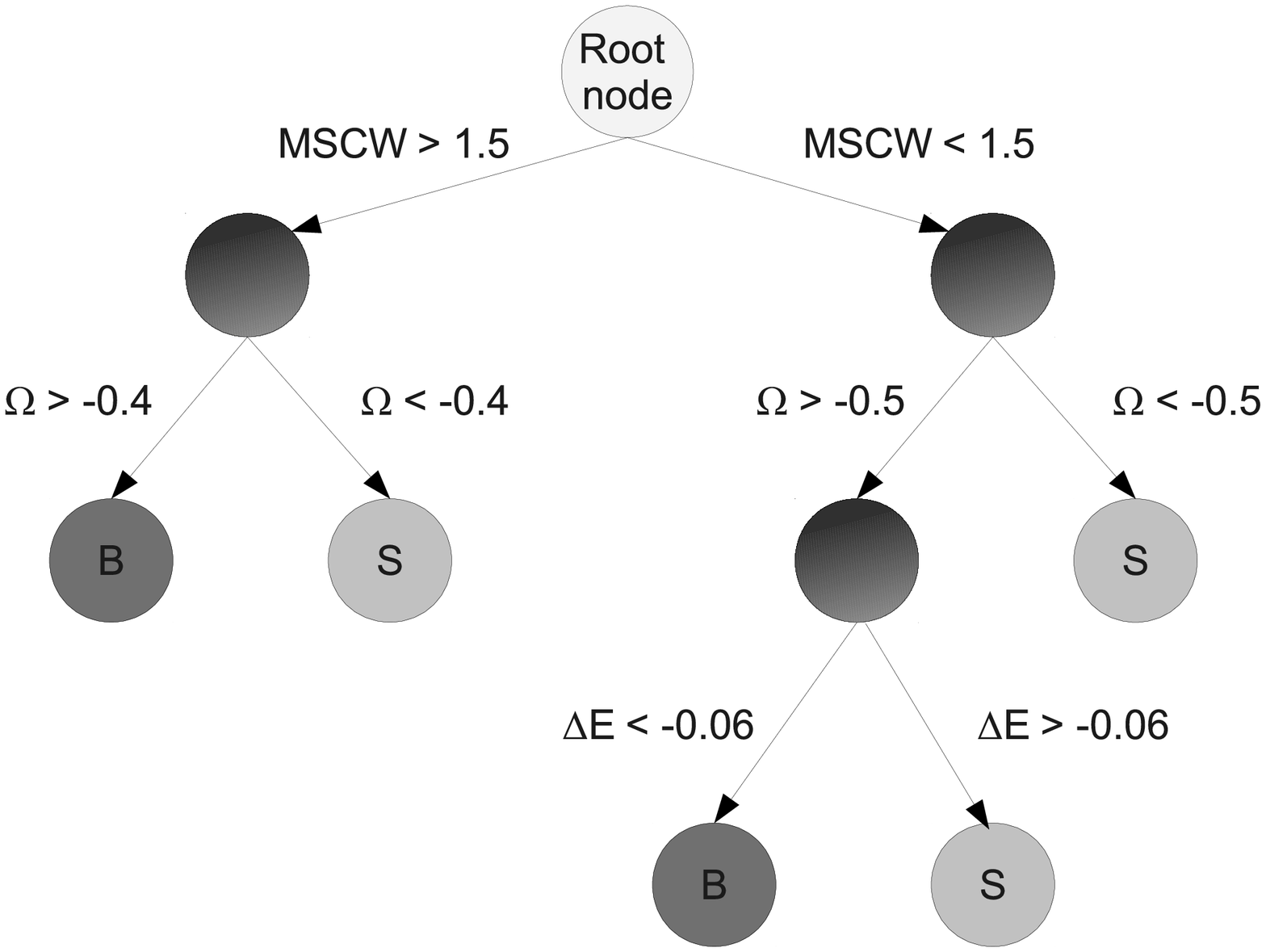}}
  {\begin{minipage}[t]{0.4\textwidth}
      \begin{small}
        \vspace{-7cm}
        \begin{tabular}{|m{3.6cm}|m{2.4cm}|}
          \hline
          \hline
          Parameter & Value \\
          \hline
          \hline
          SeparationType      & Gini index        \\  
          MaxDepth            & 20        \\  
          nEventsMin          & 30        \\
          nCuts               & 20        \\
          Max events for training & $1.2 \cdot 10^{5}$ \\
          trees/events        & 100/10000 \\
          Max events for test     & $1.2 \cdot 10^{5}$ \\
          VarTransform        & None      \\
          UseYesNoLeaf        & True      \\
          BoostType           & AdaBoost  \\
          UseWeightedTrees    & True           \\
          PruneMethod         & CostComplexity \\
          PruneStrength       & -1             \\
          NodePurityLimit     & 0.5            \\
          \hline
          \hline
        \end{tabular}
      \end{small}
  \end{minipage}}
    \caption{\small{\textsl{Left panel:}
      Schematic view of a decision tree. 
      Starting from the root node, an example of a possible
      tree development is illustrated. A sequence of binary splits using
      the defined discriminant variables is applied to the data. 
      At each node, a decision between signal and background 
      is carried out using the variable giving the
      best separation performance. See \cite{tmva} for further details.
      \textsl{Right panel:} Parameters used to construct the architecture of the Boosted Decision Trees 
      applied for the discrimination procedure presented.}}
   \label{BDTree}
\end{figure}
The comparison of the two classifiers is usually evaluated by a Kolmogorov 
test on the shape of the two resulting distributions. 
As shown in Fig.\ \ref{BDTresponse}, the Kolmogorov test value is an important check on the 
reliability of the signal-to-background classification behaviour.
If the statistical test of compatibility fails, 
the reliability of the analysis results will not be guaranteed. 

\subsection{Design of the decision tree}
\label{TreeDesign}

The result of the tuning of the values defining the tree architecture
in our specific parameter environment, is summarized in the table shown in Fig.\ \ref{BDTree}b.
The tests carried out by the authors showed that the discrimination performance was enhanced when allowing 
the tree to develop quite deeply and then cutting back the branches with an optimal \textsl{PruneStrength}.  
For this reason, in this analysis the splitting can be 
carried out up to a maximum tree depth of 20 node layers (\textsl{MaxDepth}), however, 
the sequence of event separation splits is stopped once a minimum number of 30 events  
has been reached in a node (\textsl{nEventsMin}). 
Having fixed all the other training parameters, the discrimination performance can vary substantially
depending on the value of \textsl{PruneStrength}.
Tests have shown that the optimal choice was to set this value 
to $-1$, giving the possibility to use an algorithm which tries to detect 
the optimal strength of \textsl{pruning} for the given problem. \par

Every bin in this analysis is a separate entity having its own number of signal and background events, 
and a specific approach has been used for the calculation of the statistics available for each bin
and for the choice of the number of trees composing the different forests.
Given a total number of background and signal events ($N_{\rm sig}$, $N_{\rm bkg}$) available for a specific bin,
the number of events to be used during the \textsl{training} phase 
$N_{\rm train}$ is calculated in the following way:
\begin{equation}
  N_{\rm train} = {\rm min}(n_{\rm train}, 1.2 \times 10^{5}); \;\;\; {\rm where} \;\;
  n_{\rm train} = {\rm min}(0.6 \times N_{\rm sig}, 0.6 \times N_{\rm bkg}).
  \label{ntrain}
\end{equation}
The resulting number of events for the different bins can vary considerably due to threshold 
effects as a function of the zenith angle and due to the available simulations and real observations. 
In the 112 bins, $N_{\rm train}$ can vary from $2 \cdot 10^{3}$ to $3 \cdot 10^{4}$. 
Finally, the number of events during the \textsl{test} phase is calculated by the remaining set of events after 
$N_{\rm train}$ subtraction by:
\begin{equation}
n_{\rm test} = {\rm min}(N_{\rm sig}-N_{\rm train},N_{\rm bkg}-N_{\rm train});\;\;\; 
N_{\rm test} = {\rm min}(n_{\rm test}, 1.2 \times 10^{5})
\end{equation}

During the development of this analysis method, 
it has been observed that an instability in performance occurs when the number of requested trees in the forest 
is high and the number of events present in the sample is low.
For this reason, the number of trees composing the forest has been adjusted as a function 
of the available $N_{\rm train}$ statistics present in each bin as:
\begin{equation}
n_{\rm trees} = 100\times(N_{\rm train}/10^{4}); \;\;\; {\rm with} \;
N_{\rm trees} = \rm{min}(max(n_{\rm trees},10),200); 
\end{equation}
leading to forests composed of 10 to 200 decision trees.

\begin{figure}[t]
  {\subfloat[]  {\includegraphics[width=3in]{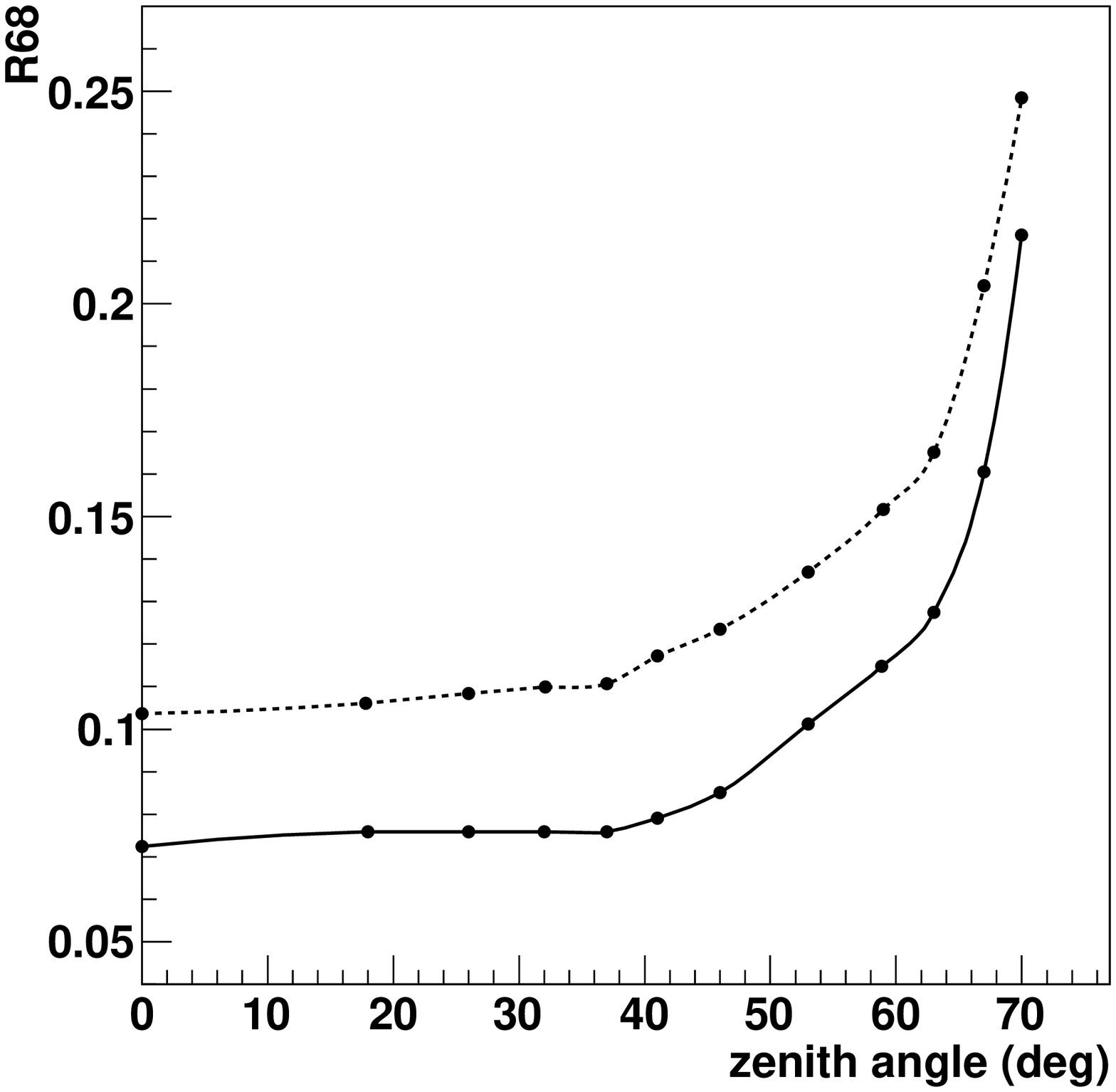}}}
  {\subfloat[]  {\includegraphics[width=3in]{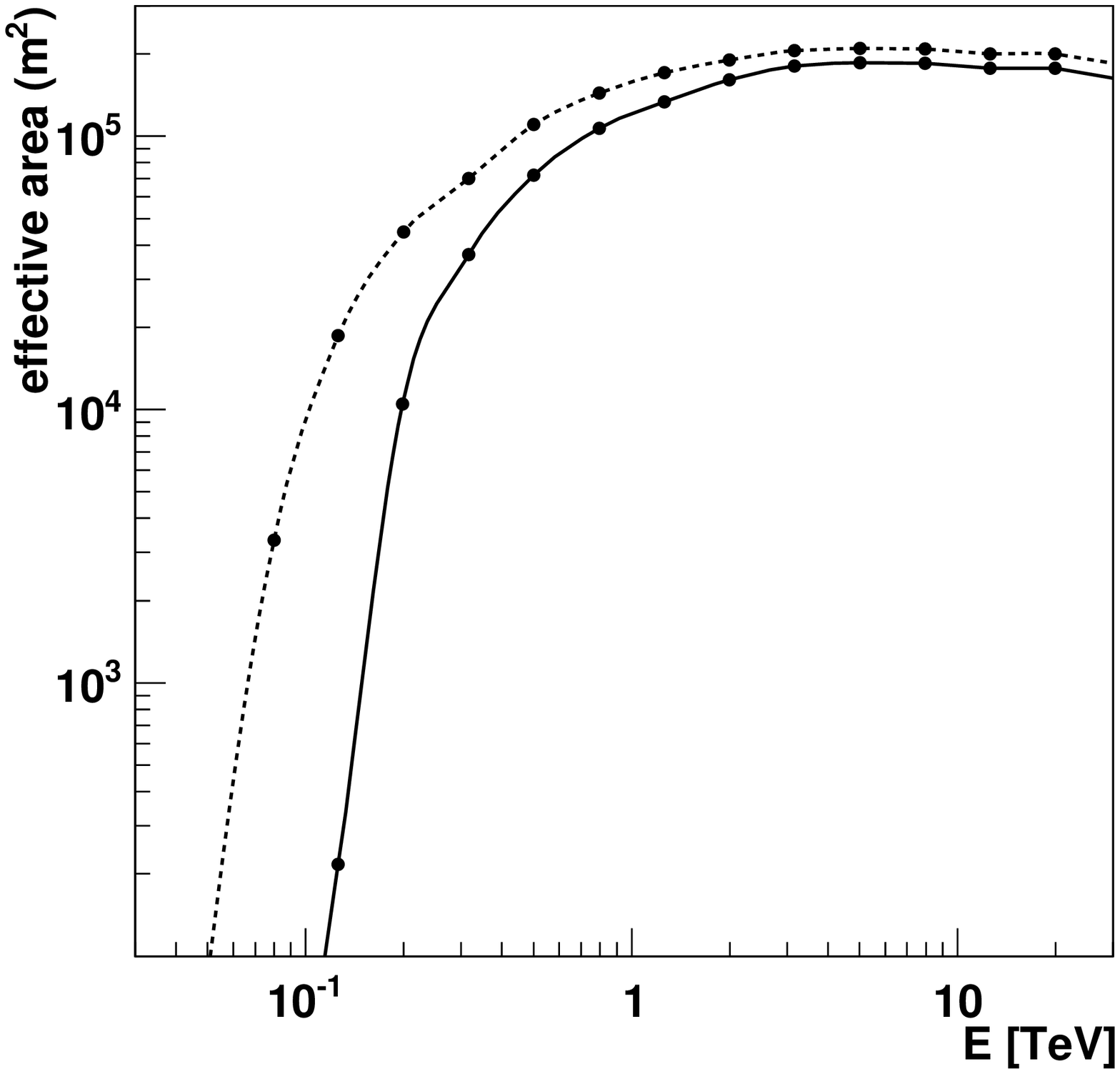}}}
  \caption{\small{
     For the two different analysis cuts
     ($40\;\rm p.e.$ and $150\;\rm p.e.$, dashed and continuous line respectively, see Sec.\ \ref{HardSoftSpectra}).
     (a) Point-spread functions defined as the containment radius at $68\%$ of all the events 
     passing the multivariate cuts as a function of the observation zenith angle for $0.5^\circ$ offset.
     (b) Effective area of events passing 
     the multivariate cut and the point-like source cut defined in Sec.\ \ref{PSFcuts} 
     for a zenith angle of $18^\circ$.
  }}
  \label{AllPSFs}
\end{figure}

\subsection{Cut application during analysis} 
\label{CutAppli}

The final step of the multivariate procedure is the application of the cuts during the data analysis 
using the BDT classifier adapted for the mean zenith angle of each run.
For each event in the run, the values 
of the discriminant parameters are evaluated, and, according 
to the number of photons in the shower as fitted by the 3D-model  
and the telescope multiplicity of the event, 
the weights and the cut value of the corresponding training bin can be 
identified straightforwardly. 
For events whose parameters correspond to a bin for which not many simulated events were available
for the training/test phases (which is the case for instance for the 
events whose energy is well below the threshold for a given zenith angle), no discrimination is possible 
and the event is classified as being background.

\section{Point-spread functions and effective areas} 
\label{PSFcuts}
The response of the instrument to \gray\ point sources, or \gray\ point-spread function (PSF), 
is usually characterised by the $68\%$ containment radius. 
It is important that this be small in order to allow the background to be better rejected.
The $68\%$ containment radii (see Fig.\ \ref{AllPSFs}a)
remain almost constant up to zenith angles of $40^\circ$, 
so for the evaluation of the final cuts for point-like sources
the mean values in the range $0^{\circ}$--$40^{\circ}$ 
for the analysis configuration under consideration are taken, and,
given that the systematic error of the pointing accuracy is $0.005^\circ$, 
slightly larger values are applied as cuts for the final analysis: 
a good compromise is found with 
$0.11^{\circ}$ at $40\;\rm  p.e.$, $0.10^{\circ}$ at $80\;\rm  p.e.$ (and $60\;\rm p.e.$) 
and finally $0.09^{\circ}$ at $110\;\rm p.e.$
The $150\;\rm p.e.$ configuration has been conceived for source morphology studies where a very 
good PSF is requested. As shown in Fig.\ \ref{AllPSFs}a, thanks to the charge threshold and 
the chosen \gray\ efficiencies, the $68\%$ containment radius for this latter configuration reaches $0.075^{\circ}$. 
The effective detection areas in the case of a point-like source 
for the analysis configurations, evaluated after this additional angular cut, are shown in Fig.\ \ref{AllPSFs}b.
As expected from the different minimal charge values for
the Hillas-parameter based reconstruction, 
the configurations have different energy thresholds and 
different integrated effective areas.

When setting the charge threshold at $40\;\rm  p.e.$, a mean gain 
in detection threshold of a factor of about $2$ 
with respect to the $150\;\rm  p.e.$ charge threshold is seen, 
as can be deduced from effective areas in Fig.\ \ref{AllPSFs}b,
which is extremely important for the study 
of sources with soft spectra.
On the other hand, when a higher charge threshold is set,
the effective area is smaller, with the compensation of a better PSF. 

\end{appendices}

\section{Acknowledgements}
The authors would like to thank Prof. W. Hofmann, spokesperson of the \hess\ Collaboration and
Prof. G. Fontaine, chair of the Collaboration board, for the permission to use the \hess\ 
data in this publication. YB would like to thank Patrick Fleury and the referees for reading 
the manuscript and for providing extremely useful suggestions.

\bibliographystyle{elsarticle-num}

\end{document}